\documentclass[10pt,twocolumn]{article}

\usepackage{lipsum} 		
\usepackage{blindtext} 		

\usepackage[superscript, biblabel]{cite}  	

\makeatletter
\renewcommand\@biblabel[1]{#1.}
\makeatother

\usepackage{etoolbox}
\patchcmd{\thebibliography}{\section*{\refname}}{}{}{}

\usepackage[margin=1.0in,hmarginratio=1:1,top=32mm,columnsep=15pt]{geometry}

\usepackage{color}				
\definecolor{dark-gray}{gray}{0.1}	
\usepackage{times}				
\usepackage{microtype} 			
\usepackage[mathscr]{euscript}  	

\usepackage{amsmath}
\usepackage{amssymb}
\usepackage[Symbol]{upgreek}
\DeclareMathAlphabet{\mathscrbf}{OMS}{mdugm}{b}{n}  
\DeclareMathSizes{10}{9}{7}{5} 

\usepackage[hang, font={footnotesize}, labelfont={bf,up}, textfont={sf,up}]{caption} 
\usepackage{booktabs} 	
\usepackage{mwe}  		
\usepackage{graphicx}
\usepackage{xcolor}
\usepackage{caption}
\usepackage{multicol}
\usepackage{multirow}
\usepackage{array}
\usepackage{longtable}   
\usepackage{morefloats}  

\usepackage{enumitem} 		
\setlist[itemize]{noitemsep} 	

\usepackage{abstract} 
\AtBeginDocument{}  		

\usepackage{titlesec} 			
\renewcommand\thesection{\Roman{section}.} 		  		
\renewcommand\thesubsection{\thesection\Alph{subsection}.} 	
\renewcommand\thesubsubsection{\thesubsection\arabic{subsubsection}.} 
\titleformat{\section}[block]{\normalfont\sffamily\bfseries}{\thesection}{1em}{\MakeUppercase}{} 	
\titleformat{\subsection}[block]{\normalfont\sffamily\bfseries}{\thesubsection}{1em}{}{}  
\titleformat{\subsubsection}[block]{\normalfont\sffamily\bfseries}{\thesubsubsection}{1em}{}{}  
\titlespacing*{\section}{0.0em}{1em}{0.25em}		
\titlespacing*{\subsection}{0.0em}{1em}{0.25em}	
\usepackage{indentfirst}						

\usepackage{pdfpages}  

\usepackage{fancyhdr} 

\fancypagestyle{plain}{
\fancyhf{}
\lhead{\color{dark-gray}\textit{}} 
\rhead{ {\it Submitted to ANS/NT (2021). LA-UR-21-20144.} }
}

\usepackage{titling} 		

\usepackage{hyperref} 	
\hypersetup{
	backref=true,       
    	pagebackref=true,               
    	hyperindex=true,                
    	colorlinks=true,                
    	breaklinks=true,                
    	urlcolor= black,                
    	linkcolor=blue,                
    	bookmarks=true,                 
    	bookmarksopen=false,
    	filecolor=black,
    	citecolor=blue
}

\title{
\vspace{-0.3in} 
\sffamily{\large\bfseries
On the origins of Lagrangian hydrodynamic methods}
\vspace{-0.1in}
}	
\author{%
\normalsize  Nathaniel R. Morgan \thanks{corresponding author: nmorgan@lanl.gov}\, and Billy J. Archer \\
\normalsize X-Computational Physics Division \\ [-0.5ex] 
\normalsize Los Alamos National Laboratory \\ [-0.5ex] 
\normalsize Los Alamos, NM 87545
}
\date{ } 
\renewcommand{\maketitlehookd}{%
\begin{abstract}
\vspace*{-0.5in}
\noindent {\normalfont\textbf{Abstract}\textemdash}
\noindent 
The intent of this paper is to discuss the history and origins of Lagrangian hydrodynamic methods for simulating shock driven flows.  The majority of the pioneering research occurred within the Manhattan Project.  A range of Lagrangian hydrodynamic schemes were created between 1943 and 1948 by John von Neumann, Rudolf Peierls, Tony Skyrme, and Robert Richtmyer. These schemes varied significantly from each other; however, they all used a staggered-grid and finite difference approximations of the derivatives in the governing equations, where the first scheme was by von Neumann. 
These ground-breaking schemes were principally published in Los Alamos laboratory reports that were eventually declassified many decades after authorship, which motivates us to document the work and describe the accompanying history in a paper that is accessible to the broader scientific community. Furthermore, we seek to correct historical omissions on the pivotal contributions made by Peierls and Skyrme to creating robust Lagrangian hydrodynamic methods for simulating shock driven flows. Understanding the history of Lagrangian hydrodynamic methods can help explain the origins of many modern schemes and may inspire the pursuit of new schemes.

\end{abstract}
}

\begin{document}

\maketitle	

\section{Prologue}
\label{sec:prolog}
We will start by providing an overview of the history surrounding the hydrodynamic methods work between 1943 and 1948 to establish context and communicate the sequence of events that occurred over that time frame to yield a suite of Lagrangian hydrodynamic methods. Subsequent sections in the paper will provide further details on the governing formulations and the four distinct numerical methods by John von Neumann, Rudolf Peierls, Tony H.R Skyrme, and then Robert Richtmyer.

Researchers at Los Alamos seemed surprisingly ready in early 1944 to numerically solve the partial differential equations for shock hydrodynamics. Looking back to 1940 through 1943, this readiness is not surprising. With the advent of World War II (WWII) the effects of the blast from bombs and artillery became of intense interest. Blast waves are described by a set of non-linear, time-dependent, partial differential equations. Further, the shock wave in an inviscid fluid has a discontinuous solution. These equations can only be solved analytically for a few very special cases, in general the solution requires numerical methods. Exploration of blast wave effects were organized in both the United States (U.S.) and the United Kingdom (U.K.). In the U.S., the National Defense Research Committee (NDRC) was established in June 1940. Section B concerned ``Bombs, Fuels, Gases, and Chemical Problems''\cite{1}. When the NDRC was superseded by the Office of Scientific Research and Development (OSRD) in June 1941, parts of the NDRC Section B became Division 2, ``Effects of Impact and Explosion'', and Division 8, ``Explosives''\cite{1}.  Vannevar Bush was the chairman of OSRD with James B. Conant as chairman of NDRC.

In May 1942 OSRD Division 8 issued a report on the ``general conditions for the existence of shock waves'' by Hans Bethe, endorsed by George B. Kistiakowsky, Chairman of Section B-1\cite{2}. Hans Bethe was also exploring the pressure wave generated by underwater explosions\cite{3,4}. von Neumann published a series of OSRD reports on high explosive detonation waves and shock waves in 1942 and 1943\cite{5, 6, 7, 8}. See the Division 2 summary for the many reports on explosion effects\cite{9}.

Meanwhile in England, William G. Penney was carrying out research on shock waves due to explosions, summarized in 1950\cite{10}, and also on the effects of blast waves during the London Blitz\cite{11}, Geoffrey I. Taylor was a member of the British delegation with long experience in hydrodynamics, shock waves, and explosives\cite{11, 12, 13, 14}. Peierls was also working on the effects of blasts, corresponding with Taylor and Kistiakowsky\cite{11}. Peierls brought his assistants  Klaus Fuchs and Skyrme to Los Alamos as members of the British delegation.  Skyrme would make significant discoveries that enabled robust and accurate hydrodynamic calculations with strong shocks. By the time Project Y, the Los Alamos branch of the Manhattan Project, was formed in 1943 there was broad experience with shock hydrodynamics by the senior staff of both the U.S. and U.K. 

During the summer of 1943 Robert Serber and Edward Teller derived an analytic approach to solve the shock hydrodynamics problem\cite{15}.
A set of IBM Punch Card Accounting Machines (PCAM) had been ordered in January 1944 for use by Stanley Frankel and Eldred Nelson to solve the gun bomb neutronics problem. 
In March 1944 the tasking of Frankel and Nelson was changed to numerically solve a range of problems, including hydrodynamics problems.
See the companion ANS article for a discussion of the PCAM and the computing facility\cite{ArcherComputers}.

In March 1944, Hans Bethe reorganized T-Division, with the IBM calculations in Group T-2 under Serber\cite{18}.
\begin{itemize}
    \item T-1, ``Hydrodynamics of Implosion, Super'', Edward Teller
    \item T-2, ``Diffusion Theory, IBM Calculations, and Experiments'', Robert Serber
    \item T-3, ``Experiments, Efficiency Calculations, Radiation Hydrodynamics'', Victor Weisskopf
    \item T-4, ``Diffusion Problems'', Richard Feynman
    \item T-5, ``Computations'', Donald Flanders
\end{itemize}
Peierls moved to Los Alamos in June 1944, replacing Teller as T-1 Group Leader.
Skyrme joined Los Alamos in T-1 in July 1944.
The stage was now set for researchers at Los Alamos to create hydrodynamic methods to simulate problems with strong shocks.

\section{History}
\label{sec:history}
After von Neumann visited Los Alamos in September and October 1943, he studied how to numerically solve the shock hydrodynamics equations. 
The first 1D Lagrangian hydrodynamic method was a staggered-grid finite difference method, created by von Neumann, and was published in an OSRD report, AMP 108.1R, in March 1944\cite{22}. This first scheme was for 1D Cartesian coordinates and has not been well discussed in the broader literature, yet it is the foundation that most Lagrangian hydrodynamic methods are built upon.  Soon thereafter von Neumann extended his method to 1D spherical coordinates, but that scheme was never published beyond Los Alamos laboratory reports\cite{31}. Von Neumann tested the method on a radially-outward propagating blast wave using the IBM PCAM at the Army Ballistics Research Laboratory (BRL) at Aberdeen Maryland in early 1944\cite{22}. Von Nuemann's early finite difference scheme suffered from oscillations at the shock front.  
At the time, there was much debate over the physical correctness and source of the oscillations\cite{31,37}. Several alternate approaches would later be pursued to address the oscillations. Section \ref{sec:vnhydro} describes von Neumann's previously unpublished spherical coordinate hydrodynamic method. 

Meanwhile Peierls visited Los Alamos in February 1944\cite{23}. In his autobiography, Peierls recounts how he had tried to solve the blast wave equation numerically: ``I tried a step-by-step method, in which you take the initial state as known at a chain of points a certain distance apart, and then determine approximately the state of affairs a certain time interval later. One can then repeat the process''\cite{11}. Peierls described his attempt to numerically solve the blast wave equations using the step-by-step method to the Los Alamos staff. The Los Alamos theoretical staff had been trying to solve the equations analytically\cite{15}. On February 14, 1944 J. Robert Oppenheimer wrote to General Groves about Peierls' visit\cite{24}. He noted that Peierls had explained the ``British methods'' for integrating the blast wave equations in great detail. Oppenheimer was ``not satisfied with the adequacy of the methods now in use in this laboratory'' for solving the shock hydrodynamics equations.

Peierls clearly did not know about the Courant stability condition\cite{CFL1928}, he re-discovered it empirically\cite{11}. The Los Alamos staff also did not know of the Courant stability condition, so at first they referred to it as the ``Peierls'' stability condition. Von Neumann clearly did know of the Courant stability condition, he discussed it in his March 1944 paper\cite{30}. By 1945 Los Alamos was referring to it as the Courant condition.  

Peierls sent a letter to Oppenheimer on March 15, 1944\cite{25} that described ``a new form of the shock wave equations which was suggested by Skyrme'', see appendix \ref{app:letters} to read the original letter. These equations are simpler than the ones Peierls described during his February 1944 visit to Los Alamos\cite{25}. Peierls approach required knowledge about the material conditions ahead of the shock, which Skyrme's method did not. 

Peierls and Skyrme\cite{31, 35} proposed coupling internal boundary conditions to the underlying hydrodynamic method to remove oscillations at a shock front.  With these shock-fitting methods, a higher-order central difference Lagrangian hydrodynamic method were used on either side of the shock, and then the discontinuous solutions on each side of the shock were coupled together with Rankine-Hugoniot jump conditions.  The first shock-fitting Lagrangian hydrodynamic method was created by Peierls\cite{25} and is briefly described by Skyrme\cite{31} and alluded to in a letter that he sent von Neumann in March 1944\cite{25}.  The Peierls scheme was viewed unfavorably at the time.  Skyrme says, ``The main disadvantage of this method ... was that a cycle (n+1) could not be completed until the positions of the points (i-1) etc., behind the shock had been found by numerical integration''\cite{31}.  

Skyrme created an alternative shock-fitting Lagrangian hydrodynamic method that accurately captured a discontinuous solution at a shock and evolved it across the mesh.  The method by Skyrme required an iterative solver to find the shock states. That scheme appeared to have similarities to the one by Peierls, but the details on the Peierls' scheme are limited in the available records. 
It is important to recognize that the shock fitting methods by Peierls and Skyrme were quite different from and predate the artificial viscosity based method by von Neumann and Richtmyer\cite{RichtmyerHydro,30}.   
By mid-March 1944, researchers at Los Alamos had at least four approaches available to solve the shock hydrodynamics equations: schemes by Serber, von Neumann, Peierls, and Skyrme.

Incorporating Rankine-Hugoniot jump conditions into a numerical scheme was revolutionary for that time period.  Beyond the work by Skyrme and Peierls, there was just one other shock fitting scheme at that time.  Howard Emmons\cite{Emmons1, Emmons2} published in 1944 a method for steady transonic flows over surfaces that arise in aeronautic applications.  Emmons created an Eulerian approach that combined a stream-function based method for the subsonic region with a separate method for supersonic region.  The method by Emmons differed significantly from the Lagrangian method by Skyrme and the one by Peierls that were designed for transient shock dynamics. Shock-fitting schemes similar to Peierls' and Skryme's were not published until many decades later, for example see the work by Maurizio Di Giacinto and Mauro Valorani\cite{DIGIACINTO198961}, the shock-fitting method implemented in the Los Alamos FRONTIER code\cite{GLIMM1984, SHARP1993, GLIMM2013}, the recent shock fitting book by Onofri and Paciorri\cite{ONOFRI2017}, or the modern moving Discontinuous Galerkin method with interface condition enforcement by Corrigan et. al. \cite{mDG-ICE-AndrewC, mDG-ICE-AndrewK}.  The schemes by Peierls and Skyrme were never published in a journal so they remained somewhat lost to time and unknown to the broader scientific community. The details on Skyrme's shock-fitting scheme were documented in multiple laboratory reports and letters during and after the Manhattan Project. These were declassified many decades after authorship and have since been made available to the public.  Details on Peierls' shock-fitting method have not been found in the available records. Skyrme's shock fitting scheme is described in Section \ref{sec:skyrmehydro}.

Frankel, Nelson, and Naomi Livesay developed a hydrodynamics computer program at Los Alamos for the soon-to-arrive IBM PCAM in March 1944.
The program was tested in March 1944 on the Marchant calculators with the assistance of Richard Feynman and Nicholas Metropolis\cite{23,26}. The hydrodynamics code being discussed had no name, it was just referred to as the IBM code or IBM problems.

The IBM PCAM arrived in Los Alamos on April 4, 1944.
A hydrodynamic calculation was started on the IBM PCAM ``within a week of arrival'' in early April 1944\cite{23,28}.
See the companion ANS article on the PCAM and the computing facility\cite{ArcherComputers}.
The calculation used von Neumann's spherical coordinate hydrodynamic method. Note that this method is not the method normally associated with von Neumann. The method that von Neumann and Richtmyer published in 1950 is based on artificial viscosity, and is quite different from the original method by von Neumann that was used in 1944\cite{30}. 

The internal boundary conditions with Skryme's approach was too complicated for the IBM PCAM and had to be performed by hand using Marchant calculators, with the results inserted into the IBM hydrodynamic calculation at the appropriate position and time.
This motivated research into alternative Lagrangian hydrodynamic methods. 

As an alternative to a shock-fitting approach, Peierls suggested to von Neumann in a March 28, 1944 letter that he should add a dissipative mechanism to his hydrodynamic scheme\cite{25} to remove the spurious oscillations near a shock.  Peierls said,
``Incidentally, have you thought at all about the following alternative way of avoiding discontinuity.  In actual fact, the shock front has a finite width because of the viscosity and thermal conductivity of the medium.  But artificially, assuming a viscosity very much larger than the actual, you can obtain instead of the discontinuity a front of a finite width, and as long as this width is small compared to the scale of the phenomenon, this should otherwise give no trouble.''\cite{25}  
Appendix \ref{app:letters} provides the original letter from Peierls.

In 1948, Richtmyer\cite{RichtmyerHydro} said, 
``For strong shocks it has been customary to interrupt the normal calculating routine at the discontinuity and perform a special calculation (``shock-fitting'') based on the Rankine-Hugoniot theory.''  
In that report, Richtmyer advocated for a new approach to simulate shocks using viscosity (real or fictitious) to smear the shock and achieve robust solutions without explicitly fitting the shock. He said, 
``It was pointed out by von Neumann that what is clearly needed is to take into account in the equations the dissipative mechanisms that operate in a physical shock to convert mechanical energy irreversibly into heat. This is the basis of the second proposal.''
With this approach, ``shocks are automatically taken care of by ... introducing a (real or fictitious) dissipation term...''  
Richtmyer\cite{RichtmyerHydro} provided an analysis to show the Rankine-Hugoniot jump conditions were satisfied using an artificial viscosity. Using artificial viscosity had the advantage of simplicity and could be readily implemented on the computers of that time. 

By 1948, Richtmyer had developed a shock capturing scheme that incorporated artificial viscosity \cite{RichtmyerHydro,49,30}.  Richtmyer added an artificial viscosity term to Peierls and Skryme's Lagrangian hydrodynamic scheme (without the shock-fitting boundary conditions) that was in terms of the initial coordinates.  He also proposed a new scheme to solve the internal energy evolution equation, which included evaluating the EOS in terms of specific volume and temperature rather than entropy\cite{RichtmyerHydro}, $p = p(v,T)$. He also presented an approach to solve the entropy evolution equation.  
The methods by Richtmyer to evolve internal energy and entropy, along with some of his derivations\cite{RichtmyerHydro}, were not discussed in the paper that von Neumann and he later published in 1950\cite{30}; likewise, that paper did not discuss the earlier work on hydrodynamic methods by von Neumann, Peierls, and Skyrme. We present the details on Richtmyer's 1948 scheme in Section \ref{sec:richtmyer}.  

The von Neumann and Richtmyer artificial viscosity scheme published in 1950 \cite{30} used a new staggered time integration method with a staggered spatial discretization that was in terms of the initial coordinates, where the latter came from the hydrodynamic scheme by Peierls and Skyrme, and the spatially staggered-grid approach came from earlier work by von Neumann \cite{22}. 
The origins of many modern Lagrangian hydrodynamic schemes can be traced to the pioneering work by Richtmyer and von Neumann, and the impact of that work will be discussed further in the Conclusions section.

\subsection{Historical documentation}
\label{sec:challenges}
Thoroughly understanding the specifics of the early Lagrangian hydrodynamic methods is somewhat challenging for a range of reasons. There are a limited set of historical reports that document the work, and these reports were not necessarily intended to educate an outside audience.  The authors were inventing the syntax to describe the numerical approaches, so at times, their syntax differs from what is used in many journal papers and textbooks; likewise, many descriptions in the report centered around using the IBM PCAM e.g., punch cards, which is quite foreign to modern-day readers. The documents also contain typographical errors.
The Los Alamos staff at the time were experimenting with various methods of handling the shocks so many options were reported. They were also pioneering the transition from classical analytic methods to numeric methods, so many of the reports discuss both. The rest of this paper seeks to decipher this complicated story of the first numerical methods to simulate shock dynamics.

\section{John von Neumann's methods}
\label{sec:vnhydro}
von Neumann derived the Lagrange equations in terms of mass coordinates for the case of compressible, non-viscous, non-conductive hydrodynamics in his March 1944 paper\cite{22}.  In that paper, the position at time $t$ is $x = x(a,t)$ where ``the substance contained in the interval $a$, $a+da$ has the mass $da$''\cite{22}. For the rest of the paper, we will use the syntax by Skyrme, Peierls, and Richtmyer for consistency purposes, which deviates from what was used by von Neumann in his paper\cite{22}. Using the alternate syntax, the Lagrange evolution equations for a position $X$ in 1D Cartesian coordinates is

\begin{equation}
\label{eq:0a}
\frac{\partial^2 X}{\partial t^2} = -\frac{\partial p}{\partial m}
\end{equation}

\noindent where the pressure is $p$ and the mass is $m$.  The current position is $X$, which deviates from today's commonly used convention that denotes variables in the initial coordinates with capital letters, and the current coordinates with lower-case letters.  The specific volume in 1D Cartesian coordinates is

\begin{equation}
\label{eq:0b}
v = \frac{\partial X}{\partial m} = \frac{1}{\rho}
\end{equation}

\noindent 
von Neumann presented a finite difference approach (based on central differences) to solve these governing equations numerically on a staggered-grid. 
This is the origin of the staggered-grid approach to solve the Lagrange hydrodynamics equations that is still in use today.  
The focus of von Neumann's original paper\cite{22} was on 1D Cartesian coordinates, but at the end of his paper, a short discussion was given on extending the scheme to spherical coordinates.  

We will now shift the focus to von Neumann's work to extend his original scheme to spherical coordinates that was documented in laboratory reports such as the one by Skyrme\cite{31}. 
Shortly after arriving at Los Alamos, Skyrme adapted his shock-fitting treatment, mentioned in Peierls' March 15, 1944 letter\cite{25}, to work with von Neumann's hydrodynamic method\cite{22}, and he documented the method in LAMS-562\cite{31}. 
In what follows, we will adopt the notation of LAMS-562\cite{31}.

The Lagrange evolution equation for the radius $R$ in 1D spherical coordinates is,

\begin{equation}
\frac{\partial^2 R}{\partial t^2} = -3R^2 \frac{\partial p}{\partial m}
\end{equation}

\noindent Here $m$ is the mass coordinate of an individual Lagrangian mass-point, ``equal to ($\frac{3\pi}{4}$) times the mass enclosed within the spherical shell corresponding to any point''\cite{31}, The position of a mass element at a particular time, t, is defined as $R$. The specific volume, $v$, is defined as

\begin{equation}
\label{eq:1}
v = \frac{\partial R^3}{\partial m} = \frac{1}{\rho}
\end{equation}

von Neumann presented a staggered-grid finite difference method to solve the governing evolution equations.  If the time cycle is labeled by a superscript $n$, and the mass-points are labeled by a subscript $i$, the discrete equation to be solved is 

\begin{equation}
R^{n+1}_i = 2R^n_i - R^{n-1}_i - 3(R^n_i)^2\,(\Delta t)^2 \, \frac{p^n_{i+\frac{1}{2}} - p^n_{i-\frac{1}{2}} }{\Delta m}
\end{equation}

\begin{figure} 
\centering
\includegraphics[trim=0 0 0 1.1in, width=2.25in,height=!,clip=true]
{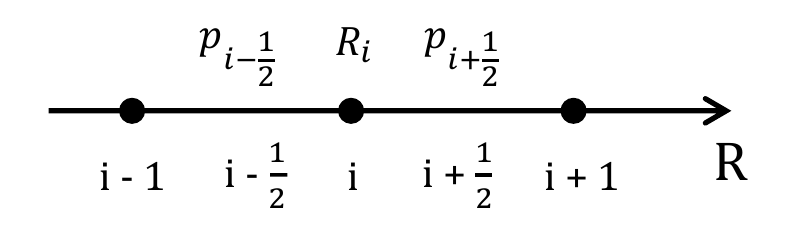}
\caption{\label{fig:mesh_variables_vn} The staggered mesh used with von Neumann's first Lagrangian hydrodynamic method}
\end{figure}

\noindent See Figure \ref{fig:mesh_variables_vn} for a graphical illustration of the variable locations. 

\noindent The time difference is

\begin{equation}
\Delta t = t^{n+1} - t^n =  t^{n} - t^{n-1}
\end{equation}

\noindent The mass difference is

\begin{equation}
\Delta m = m_{i+\frac{1}{2}} - m_{i-\frac{1}{2}} = \frac{m_{i+1} - m_{i}}{2}
\end{equation}

\noindent A key assumption that von Neumann made was that entropy was constant everywhere away from  the shocks. The pressure is determined using the equation-of-state from the entropy $S$ and the specific volume

\begin{equation}
p^n_{i+\frac{1}{2}} = p(v^n_{i+\frac{1}{2}}, S)
\end{equation}

\noindent The discrete specific volume is

\begin{equation}
v^n_{i+\frac{1}{2}} = \frac{ (R^n_{i+1})^3 - (R^n_{i})^3 }{m_{i+1} - m_{i}}
\end{equation}

\noindent Because the pressure could be calculated from the entropy and specific volume there was no need to solve the energy equation. The pressure was calculated at the mid-points between the mass-points, this was a staggered-grid Lagrangian formulation. 

von Neumann pointed out the Courant time stability condition\cite{CFL1928, 22}. Peierls had empirically found the same condition\cite{11}.
Skyrme terms it the 'L-condition' in his report\cite{31}. Defining the local sound speed,

\begin{equation}
c = \sqrt{ \frac{\partial p}{\partial \rho} \bigg|_{S} }  
\end{equation}

\noindent were the partial derivative is taken with entropy $S$ remaining constant, and should not be confused with the subscript $s$ that is used later in the paper to denote a quantity at a shock. The Courant limit is

\begin{equation}
L = \bigg|\frac{c\Delta t}{\Delta R}\bigg| = 
    \bigg|\frac{3 R^2 \rho c\Delta t}{\Delta m}\bigg| < 1
\end{equation}

\noindent A value of $L=1/2$ is reported to be used\cite{31}. As noted above, the equation of state (EOS) is needed to solve the difference equation. 
The final form of the EOS was determined so as to limit the number of punch cards required, from LA-1058\cite{35}, 

\begin{equation}
p(v,S) = F_1\frac{v}{a} + bF_2\frac{v}{a}
\end{equation}

\noindent Here $F_1$ and $F_2$ are constants for a material, $v$ is the specific volume, and the coefficients $a$ and $b$ are known functions of $S$, the entropy. The entropy was calculated from the specific volume\cite{35}. One thousand punch cards\cite{35} were used to tabulate all possible values of $v/a$ and $b$. 
This completes the hydrodynamics equations that were used with von Neumann's 1D spherical coordinate method.

von Neumann proposed using this method for treating shocks by just letting the hydro deal with it, i.e., ``completely ignoring the possibility of shocks''\cite{22}. He visualized the problem as a set of mass points connected by springs. He believed that the above hydrodynamic equations would allow good approximations of a shock, but with oscillations behind the shock. ``As soon as a shock has been crossed such oscillations must develop, and they have a perfectly good physical significance. They represent the thermic agitation caused by the degradation of energy through the shock''\cite{22}. Only the average hydrodynamic values were valid behind the shock. The test problems von Neumann carried out on the BRL PCAM in early 1944 were air blasts that had weak shocks\cite{22}. The approach of letting an inviscid hydrodynamic method calculate the shocks, especially weak shocks will be referred to as the von Neumann method in this paper. 
When tried on strong shocks, von Neumann's method created more violent oscillations than expected. Peierls showed that von Neumann's method was only valid for weak shocks\cite{31,37}. 
The conclusion was that von Neumann's method could not be used for strong shocks, and lead to them using shock-fitting methods.

\section{Rudolf Peierls' Shock-fitting Method}
\label{sec:peierlshydro}
Peierls proposed an approach in his March 28, 1944 letter to von Neumann that treated the ``shock wave classically''\cite{25,31}.
This method required extrapolating the shock position to the next time step\cite{25}. Metropolis worked out the numerics of Peierls shock method in April 1944. 
Skyrme described Peierls' shock fitting method in LAMS-562\cite{31}. Skyrme stated that the ``main disadvantage of this method, apart from the fact that it introduced undesirable fluctuations behind the shock, was that a cycle (n+1) could not be completed until the positions of the points (i-1) etc. behind the shock had been found by numerical integration. This introduced a delay between cycles''\cite{31}. This is the same weakness as the extrapolation method Peierls described in his March 28, 1944 letter\cite{25}.

For strong shock problems, the pressure found in the ``classical calculation'' with Peierls' method agreed closely with the average pressure calculated using von Neumann's method. 
However, the classical method showed that the high pressures from von Neumann's method on strong shock problems were spurious.

\section{Tony Skyrme's Shock-Fitting Method}
\label{sec:skyrmehydro}
Neither von Neumann's nor Peierls' shock methods were really satisfactory. Peierls' classical method required information ahead of the shock, and von Neumann's method was only valid for weak shocks\cite{31, 37}. This, coupled with the observation of the constant in space pressure behind the shock, lead to the use of a third method for handling strong shocks that also reduced the ``undesirable fluctuations behind the shock''.

Skyrme dramatically increased the sophistication of handling shocks with the intent of doing the best possible simulation. The shock-fitting method is documented in LAMS-562\cite{31} and in LA-1058\cite{35}. LAMS-562 was published in 1947 after Skyrme left Los Alamos, but it appears to have been written in December 1944 or January 1945.  Similarly, LA-1058 was published in June 1949, but it contains a chapter written by Skyrme that thoroughly described his shock-fitting hydrodynamic method.
  
In both LAMS-562 and LA-1058, Skyrme described a novel iterative approach to impose the Rankine-Hugoniot jump conditions at the shock front within a Lagrangian hydrodynamic calculation.  
Skyrme also presented a new Lagrangian staggered-grid hydrodynamic method based on the initial coordinates that is also referred to in an early letter by Peierls in 1944\cite{25}. Skyrme's hydrodynamic scheme was applied everywhere on the mesh in LA-1058, whereas in the case of LAMS-562, his hydrodynamic method was only used near the shock and von Neumann's method was used everywhere else.\cite{31}

For strong shocks, Peierls' shock fitting method was replaced with the shock fitting approach of Skyrme\cite{31}. 
Skyrme's method shown in LAMS-562\cite{31} has 63 steps! The ``procedure used in computing the position of the shock is so complicated that it is impractical to use the I.B.M. machines for it, but it can be done satisfactorily with an ordinary calculating machine''\cite{35}. Even so, the shock procedure on the Marchant for a time step could be completed in about the same time it took the IBM PCAM to calculate a single hydro cycle\cite{35, 39}.

By March 1945 a standard shock procedure was in place, Skyrme's shock-fitting approach, 
combined with von Neumann's hydrodynamic method, 
was used for the strong shocks\cite{31}. 
The weak shocks were treated using only von Neumann's method\cite{31}.    
By the late 1940's, researchers had switched to using Skyrme's Lagrangian hydrodynamic method (which is in terms of the initial coordinates) in place of von Neumann's method (which used mass coordinates) but still with the shock-fitting approach.

The remainder of this section will focus on describing the key equations used with the shock-fitting scheme by Skryme.  We will present the key parts for the scheme documented in LA-1058 as it is the most complete. There appears to be only one minor change between the scheme documented in LAMS-562 and LA-1058, which will be highlighted. The method in LA-1058 assumes an initial mesh that is at rest with a single outward propagating shock starting at the origin (akin to the Sedov blast wave problem\cite{Sedov, MorganSedov}); the presentation that follows adheres to this assumption.  As a note of caution to the reader, the original manuscripts by Skyrme contained typographical errors such as placing a minus sign on an equation, omitting a superscript on a variable, or not including a square root sign on a term. As such, the equations presented in this paper will in a few cases deviate slightly from what was printed in the original manuscripts.

\subsection{Governing Lagrangian hydrodynamic equations}
\label{sec:skyrme-gov-eqns}
The shock fitting hydrodynamic methods by both Peierls and Skyrme solved the governing analytic Lagrangian hydrodynamic equations in terms of the position in the initial spatial coordinates $r$, which are termed total Lagrangian methods in modern literature.  The position in current spatial coordinates is $R$.  We follow the naming convention used by Peierls and Skyrme.  The specific volume of the material in the initial coordinates is $v_0$ and the pressure is $p$.  The evolution of a position in the current spatial coordinates is,

\begin{equation} 
\label{mom}
 	\frac{d^2 R}{d t^2}= - v_0  \frac{ R^2 }{r^2} \frac{dp}{dr} 
\end{equation}

\noindent The specific volume at time $t$ is

\begin{equation} 
\label{specVol}
 	v = v_0  \frac{ \partial R^3 }{\partial r^3} 
\end{equation}

\noindent It is reasonable to assume that the flow is adiabatic i.e., there is no time for heat to transfer away.  A continuous flow will be isentropic.  The only change in entropy is across the shock; as such, the rate of change of the entropy with respect to time away from a shock is zero.

\subsection{Rankine-Hugoniot jump relations} 

This subsection presents the Rankine-Hugoniot relations used in the scheme by Skyrme\cite{31, 35}.  The velocity of the shock is

\begin{equation} 
\label{dr_sdt}
 	\frac{d r_s}{d t} = v_0 \sqrt{ \frac{p_s - p_0}{v_0 - v_s} } 
\end{equation}

\noindent The pressure at the shock is $p_s$ and the specific volume of the shock is $v_s$.  The shock pressure is a function of the specific volume and entropy, $p_s = \psi(v_s, S_s)$.   The material velocity at the shock is

\begin{equation} 
	u_s \equiv \left( \frac{\partial R}{\partial t}\right)_s = \sqrt{ (p_s - p_0)(v_0 - v_s) }
\end{equation}

\noindent The acceleration of the shock is

\begin{equation}
\label{d2r_sdt2}
 	\frac{d^2 r_s}{d t^2} = \frac{v_0}{2u_s}\frac{d v_s}{d t} (W^2 - w^2)
\end{equation}

\noindent where 

\begin{equation}
	w \equiv \sqrt{ -\frac{\partial p}{\partial v}\,} \quad \text{and} \quad W \equiv \sqrt{ \frac{p_s - p_0}{v_0 - v_s} } 
\end{equation}

\noindent The acoustic impedance is $\rho c =  \sqrt{-\frac{\partial p}{\partial v}\big|_S}$ and is a material dependent variable that was tabulated.  The rate of change of the specific volume at the shock is

\begin{multline}
\label{dv_sdt}
 	\frac{d v_s}{d t} =  \\
	\frac{2W}{3W^2 + w^2}
	\left({ \frac{2 u_s v_s W}{R_s} 
	 + v_0 W^{2}\left(\frac{\partial v}{\partial r}\right)_s  
	 + v_0 \left( \frac{\partial p}{\partial r}\right)_s
	}\right)
\end{multline}	

\noindent where $R_s$ is the spherical radius at the shock front in the current coordinates. This equation is a function of the state of the material that is being shocked $(p_0, v_0)$ and a function of $v_s$. As a reminder, all presented equations assume a shock traveling in the positive radial direction into a medium that is at rest.

\subsection{Finite difference solver for shocks}

The specific volume at time $n$ is calculated using, 

\begin{equation}
v^{n}_{i - \frac{1}{2}} = 
	v_o \frac{ (R_i^n)^3 - (R_{i-1}^n)^3}{(r_i^n)^3 - (r_{i-1}^n)^3}
\end{equation}

\begin{figure*} 
\centering
\includegraphics[trim=0 0 0 0in, width=4.5in,height=!,clip=true]
{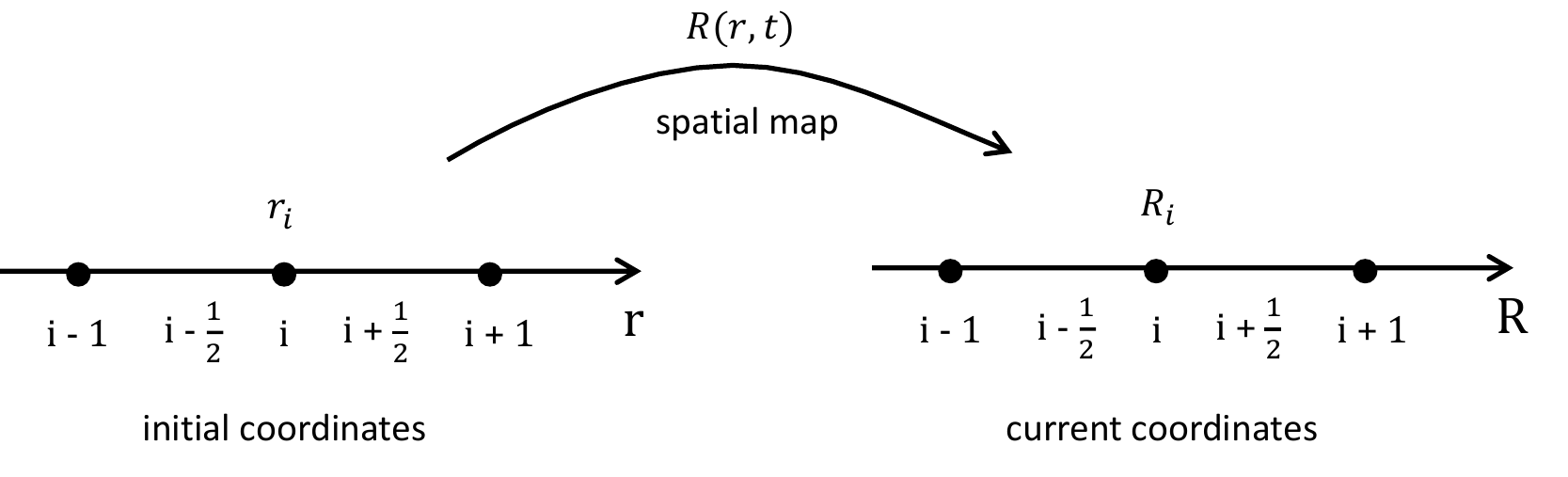}
\caption{\label{fig:mesh_indicies} Skyrme's shock-fitting hydrodynamic method solved the governing equations in the initial coordinate system.}
\end{figure*}

\noindent This equation is the identical to the strong mass conservation used in modern methods.  The nodal positions in the initial coordinates are $r_i$ and the nodal positions in the current coordinates are $R_i$, see Figure \ref{fig:mesh_indicies} for a diagram illustrating the mesh nomenclature.
The finite difference equation for evolving nodal positions in the current coordinates away from the shock is 

\begin{equation}
\label{eqn:R_Skyrme}
R^{n+1}_i = 
	2 R^n_i
	-R^{n-1}_i
	- ({\Delta t^2}) \frac{ v_0 (R_i^n)^2 }{r^2_i} \left(\frac{ p^n_{i + \frac{1}{2}} - p^n_{i - \frac{1}{2}} }{r^n_{i + \frac{1}{2}} - r^n_{i - \frac{1}{2}}} \right)
\end{equation}

\begin{figure}
\centering
\includegraphics[trim=0 0 0 0in, width=2.25in,height=!,clip=true]
{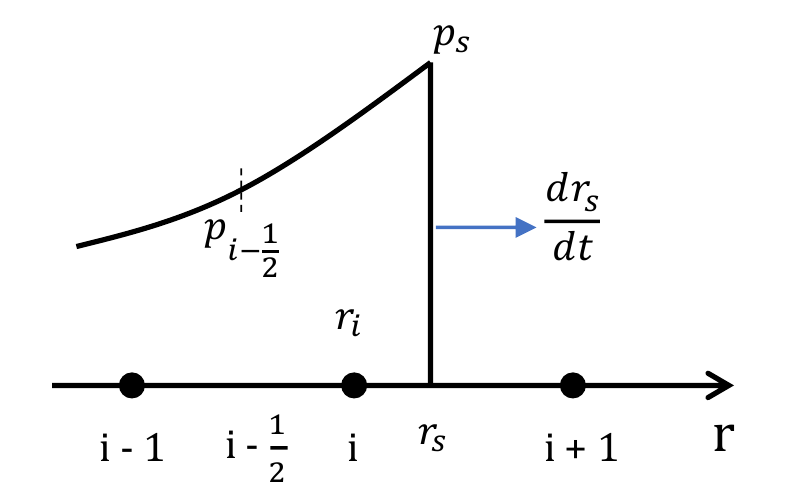}
\caption{\label{fig:mesh_variables_skyrme} A notional shock profile is shown with corresponding variables used with Skyrme's shock-fitting method}
\end{figure}

\noindent The pressure $p^n_{i - \frac{1}{2}}$ is calculated using an equation of state that is a function of $v^n_{i - \frac{1}{2}}$ and entropy $S$, which is constant away from a shock.  See Figure \ref{fig:mesh_variables_skyrme} for a diagram showing the variable locations with a notional shock profile.  

The finite difference equation shown in Eq. \ref{eqn:R_Skyrme} will give rise to large oscillations near strong shocks.  To correct these oscillations, Skyrme proposed a shock-fitting approach that introduces a discontinuity with boundary conditions to the finite difference equation.  At a shock, the pressure gradient behind the shock is based on the shock pressure instead of the neighboring pressure at $p_{i + \frac{1}{2}}$.

\begin{equation}
R^{n+1}_i = 
	2 R^n_i
	-R^{n-1}_i
	- ({\Delta t^2}) \frac{ v_0 (R_i^n)^2 }{r^2_i} \left(\frac{ p^n_{s} - p^n_{i - \frac{1}{2}} }{r^n_{s} - r^n_{i - \frac{1}{2}}} \right)
\end{equation}

\noindent As mentioned earlier, the shock is moving in the positive direction relative to node $R_i$.  The shock pressure $p^n_s$ is a function of $v^n_s$ through an equation of state.  The shock pressure and the specific volume for the shock are unknowns and must be calculated using an iterative method and the Rankine-Hugoniot jump equations.  The entropy only changes  across the shock, so the flow is isentropic away from the shock.  

\subsection{Solving the jump equations}
This subsection discusses how to calculate $v^n_s$, which in turn is used to calculate $p^n_s$ through an equation of state.  The shock pressure $p^n_s$ is used to evolve the position of the nodes $R_i $ in the current coordinate system and the shock position in the initial coordinates $r_s$.   The goal is to calculate the specific volume of the shock $v^n_s$ iteratively using

\begin{equation}
\label{vs-eqn}
v^n_s = v^{n-1}_s + \frac{\Delta t}{2} \left[ \left(\frac{d v_s}{d t} \right)^n +  \left(\frac{d v_s}{d t} \right)^{n-1} \right]
\end{equation}

\noindent where $\left(\frac{d v_s}{d t} \right)^n$, see Eq. \ref{dv_sdt}, is a function of the unshocked state  $(p_0^n, v_0^n)$, a function of the current shock position $r^n_s$ in the initial coordinates, a function of $v^n_s$, and a function of spatial derivatives at the shock $\left(\frac{\partial v}{\partial r}\right)^n_s$ and $ \left( \frac{\partial p}{\partial r}\right)^n_s$.  These latter spatial derivatives are calculated using,

\begin{equation}
\left(\frac{\partial v}{\partial r}\right)_s = \frac{ v_s - v_{i - \frac{1}{2}} }{ r_s - r_{i - \frac{1}{2}}  } 
\,\,\,\, \text{and} \,\,\,\,
\left(\frac{\partial p}{\partial r}\right)_s = \frac{ p_s - p_{i - \frac{1}{2}} }{ r_s - r_{i - \frac{1}{2}}  } 
\end{equation}

\noindent 
Eq. \ref{vs-eqn} is iteratively solved until $v^n_s$ on the left side is equal to the $v^{n}_s$ that is used to calculate $\left(\frac{d v_s}{d t} \right)^n$ on the right side. 

The shock position at the next time step is calculated using,

\begin{equation}
\label{rs_eqn}
r^{n+1}_s = r^n_s 
	+ \frac{\Delta t}{2} \left[ \left(\frac{d r_s}{d t} \right)^n +  \left(\frac{d r_s}{d t} \right)^{n-1} \right]
	+ \frac{\Delta t^2}{2} \left(\frac{d^2 r_s}{d t^2} \right)^n
\end{equation}

\noindent The first derivative of the shock position with respect to time $\left(\frac{d r_s}{d t} \right)^n$, see Eq. \ref{dr_sdt}, is a function of $v_s$.  The second derivative of the shock position with respect to time $\left(\frac{d^2 r_s}{d t^2} \right)^n$, see Eq. \ref{d2r_sdt2} is a function of $v_s$ and $\frac{dv_s}{dt}$.  Skyrme found that calculating $r^{n+1}_s$ using 

\begin{equation}
\label{rs_eqn2}
r^{n+1}_s = 2r^n_s - r^{n-1}_s 
	+ \frac{\Delta t^2}{2} \left(\frac{d^2 r_s}{d t^2} \right)^n
\end{equation}
\noindent generated oscillations, so he used Eq. \ref{rs_eqn}.

Skyrme's shock-fitting approach described in LAMS-562 is very close to what was presented in this section.  One minor difference worth noting between the two reports (LA-1058 and LA-562) is that the shock position is calculated in LAMS-562 neglecting the second derivative term in Eq. \ref{rs_eqn}.

\section{Robert Richtmyer's hydrodynamic scheme}
\label{sec:richtmyer}
An interesting historical point is at the end of Peierls' March 28, 1944 letter to von Neumann where he suggested that an artificial viscosity could be used to slightly smear the shock discontinuity, making it numerically tractable\cite{25}. Richtmyer implemented this idea in 1948\cite{RichtmyerHydro, 49}. He and von Neumann made this the standard method for handling shocks in Lagrangian hydrodynamic codes with their 1950 journal paper\cite{30}. This is known as a smeared shock method or shock capturing method. Use of artificial viscosity allowed the intrinsic treatment of strong shocks without undue smearing of the weak shocks. The details on Richtmyer's scheme will now be discussed.

The staggered-grid hydrodynamic scheme by Richtmyer\cite{RichtmyerHydro} solves the governing equations in 1D Cartesian coordinates in terms of the initial spatial coordinates $x$. His approach builds on the one by Skyrme (subsection \ref{sec:skyrme-gov-eqns}), and includes an artificial viscosity term and solves the specific internal energy evolution equation.  This is the first recorded instance of solving the specific internal energy evolution equation with a hydrodynamics method, which is now the standard practice.  The pressure is calculated using the specific volume $v$ and temperature $T$, so $p = p(v,T)$.  The prior hydrodynamic methods calculated the pressure in terms of the entropy $S$. The evolution equations for the position in the current coordinate system $X$ (in 1D Cartesian coordinates) and specific internal energy $\epsilon$ are as follows,

\begin{equation} 
\label{mom-rich}
  \frac{1}{v_0} \frac{d^2 X}{d t^2}= 
    -\frac{\partial p}{\partial x} +
 	2\alpha \bigg| \frac{\partial v}{\partial t} \bigg|
 	\frac{\partial^2 v}{\partial x \partial t}
\end{equation}

\begin{equation} 
\label{spec-eng-rich}
 	\frac{d\epsilon}{d t} = 
 	 -p\frac{\partial v}{\partial t} +
 	\alpha \bigg| \frac{\partial v}{\partial t} \bigg|^3
\end{equation}

\noindent The artificial viscosity coefficient $\alpha$ is derived based on studying the Rankine-Hugoniot jump relations.

\begin{equation}
    \alpha = \frac{\gamma+1}{2} \frac{l_s^2}{v_0^2v_s}
\end{equation}

\noindent where $l_s$ is the shock thickness and $v_s$ is the specific volume of shocked state.  As an approximation, Richtmyer proposed to use the non-shocked value of specific volume $v_0$ in place of $v_s$.  
The specific volume $v$ is calculated using

\begin{equation} 
\label{spec-vol-rich}
 	v = v_0
 	 \frac{\partial X}{\partial x}
\end{equation}

\noindent
These governing equations are discretized using a staggered arrangement and central difference approximations of the derivatives.

\begin{multline}
  \frac{1}{v_0} \frac{X^{n+1}_i - 2X^n_i + X^{n-1}_i}{(\Delta t)^2} =
    -\frac{ p^n_{i+\frac{1}{2}} - p^n_{i-\frac{1}{2}} }{\Delta x} + \\
 	2\alpha \bigg| \frac{  v^n_{i+\frac{1}{2}} + v^n_{i-\frac{1}{2}} - v^{n-1}_{i+\frac{1}{2}} - v^{n-1}_{i-\frac{1}{2}}}{2 \Delta t}\bigg|
 	\\
	\left(
 	\frac{v^{n}_{i+\frac{1}{2}} - v^{n-1}_{i+\frac{1}{2}} - v^{n}_{i-\frac{1}{2}} + v^{n-1}_{i-\frac{1}{2}}}{\Delta x \Delta t}
	\right)
\end{multline}

\begin{equation} 
 	\frac{\epsilon^{n+1}_{i+\frac{1}{2}} - \epsilon^{n}_{i+\frac{1}{2}}}{\Delta t} = 
 	 -p^n_{i+\frac{1}{2}}\frac{v^{n+1}_{i+\frac{1}{2}} - v^n_{i+\frac{1}{2}}}{\Delta t} +
 	\alpha \bigg| \frac{v^{n+1}_{i+\frac{1}{2}} - v^n_{i+\frac{1}{2}}}{\Delta t} \bigg|^3_{_{\,}}
\end{equation}

\begin{equation} 
 	v^{n+1}_{i+\frac{1}{2}} = v_0
 	 \frac{X^{n+1}_{i+\frac{1}{2}} - X^{n+1}_{i+\frac{1}{2}}}{\Delta x}
\end{equation}

\noindent The pressure is calculated using 
$p^n_{i+\frac{1}{2}} = p(v^n_{i+\frac{1}{2}}, T^n_{i+\frac{1}{2}})$.  Richtmyer also presented an alternate entropy-based approach that evolves specific entropy $s$ in place of the specific internal energy.

\begin{equation}
    T^n_{i+\frac{1}{2}} \frac{s^{n+1}_{i+\frac{1}{2}} - s^{n}_{i+\frac{1}{2}}}{\Delta T} = \alpha \bigg| \frac{v^{n+1}_{i+\frac{1}{2}} - v^n_{i+\frac{1}{2}}}{\Delta t} \bigg|^3 .
\end{equation}

In 1953, a code was implemented on the IBM Model II CPC\cite{50}. It employed both the method by von Neumann and Richtmyer that used an artificial viscosity for shocks, and also Skyrme's shock-fitting method, but some hand fitting was still required\cite{50}. By the mid-1950's, the shock fitting method of Skyrme was fully automated, and the successor code had the option to use either shock-fitting or smeared shock methods for simulating strong shocks.

\section{Conclusion}
The intent of this paper was to inform the broader community on the pioneering work by von Neumann, Peierls, Skyrme, and Richtmyer to create robust hydrodynamic methods to simulate shock drive flows. These researchers developed multiple Lagrangian hydrodynamic schemes between 1943 and 1948, and most of these schemes were never published beyond laboratory reports.  These early foundational advances led to subsequent, important computational hydrodynamic methods created by researchers at Lawerence Livermore, Los Alamos, and Sandia National Laboratories, and at laboratories and universities in the UK, France, and other countries. 

We discussed the original, 1D Cartesian coordinate staggered-grid Lagrangian scheme by von Neumann that was published in 1944, but garnered little attention by the broader community after it was published.  That 1D scheme used central difference approximations for the derivatives.  Soon thereafter, von Neumann proposed a 1D spherical coordinate Lagrangian hydrodynamic method, based on his earlier 1D scheme, to simulate a radially-outward propagating shock wave (akin to the Sedov blast wave problem) on the BRL IBM PCAM machine.  We believe this was the first shock hydrodynamics code, but that scheme was prone to spurious oscillations near a strong shock, often with amplitudes twice the mean pressure. von Neumann's original method was only suitable for weak shocks.  

At Los Alamos, von Neumann's hydrodynamic method was implemented in a computer program by Frankel, Nelson, and Livesey, who also directed a cadre of IBM PCAM operators. These efforts resulted in the Los Alamos IBM code, which would become the first hydrodynamics code to accurately and robustly simulating strong shocks.

The spurious oscillations in von Neumann's method motivated Peierls and Skyrme to investigate shock-fitting methods. We presented the details on a novel, essentially unknown shock-fitting Lagrangian hydrodynamic method created by Skyrme in 1944. It explicitly represented a shock discontinuity on a discrete mesh and coupled together two separate higher-order solutions with the Rankine-Hugoniot jump conditions. With this approach, higher-order accuracy is achieved without spurious oscillations on problems with discontinuities.  The scheme by Skyrme was an improvement over an earlier shock-fitting scheme by Peierls. 
The challenge with Skyrme's shock-fitting method was that it required an iterative solve to impose the shock boundary conditions, so it was not well suited for the IBM PCAM. Instead, the shock fitting was performed by hand using Marchant calculators with the results inserted into the IBM hydrodynamic calculation at the appropriate position and time.  The scheme by Skyrme was fully automated on a computer by the mid 1950's.

The shock-fitting approaches by Peierls and Skyrme were used with a staggered-grid Lagrangian hydrodynamic method that was in terms of the initial coordinates.  That formulation quickly replaced Von Neumann's mass coordinate formulation.  Since that time, a range of Lagrangian hydrodynamic schemes have been proposed that use the initial coordinates (or fixed reference coordinates), examples include the discontinuous Galerkin hydrodynamic method\cite{VilarDGlagcaf2012, VilarDG, LiuDGlagcaf2017, LiuDGlagSMS2019, LiuDGlagRZ2018, LiebermanStrength1D2017, LIEBERMAN2019467, DGHEburn}, the finite element hydrodynamic method\cite{BLAST}, and the residual distribution hydrodynamic method\cite{SvetlanaRD, ANN-ShockDetector}.  
Skyrme also introduced a strong mass conservation equation that has been used in many subsequent schemes\cite{Kolsky2D, BensonReview1992, BLAST, LiuDGlagcaf2017}. 

After WWII, researchers at Los Alamos were motivated to create a robust Lagrangian hydrodynamic scheme to simulate strong shocks and that could be fully run on the computers at that time.  The alternative to the Skyrme shock-fitting scheme was a shock capturing scheme that used artificial viscosity to smear the shock over several cells. 
The use of artificial viscosity for shock capturing was first suggested in a letter by Peierls to von Neumann on March 28, 1944\cite{25}. Richtmyer developed the quadratic formulation of artificial viscosity in 1948, calling it a ``fictitious'' addition to the hydrodynamics equations\cite{RichtmyerHydro, 49}.  Richtmyer added the artificial viscosity to Peierls' and Skyrme's finite difference scheme that was in terms of the initial coordinates. 
Likewise, he introduced the now standard practice of solving the internal energy evolution equation, which included evaluating the EOS as a function of specific volume and temperature\cite{RichtmyerHydro}, whereas prior works used specific volume and entropy.

The artificial viscosity approach was introduced to the broader hydrodynamics community, using Peierls' artificial viscosity nomenclature, in the well known 1950 paper by von Neumann and Richtmyer\cite{30}. That paper presented a new staggered time integration method with a spatially staggered discretization that was in terms of the initial coordinates. Peierls and Skyrme were the first to create a Lagrangian hydrodynamic scheme in terms of the initial coordinates following the staggered-grid approach from von Neumann in 1944\cite{22}.
The staggered-grid discretization combined with artificial viscosity is the basis of most Lagrangian hydrodynamic schemes\cite{WilkinsSGH, BensonReview1992, Zukas2004, BurtonFLAG2007, ALE3D2017, BurtonSGH, Caramana}. 
In 1955 at Los Alamos, Rolf Landshoff added a linear term to the artificial viscosity\cite{Landshoff1955, MorganSGHdisp}, and Marshall Rosenbluth suggested turning off the artificial viscosity during expansion\cite{RichtmyerMorton1967}. These additions to artificial viscosity were first publicly discussed in the 1957 edition of Richtmyer and Morton\cite{RichtmyerMorton1967}. Peter Lax and Burt Wendroff put the shock capturing method into conservative form in the 1960s\cite{Lax1954, LaxWendroff1958, LaxWendroff1960, RichtmyerMorton1967, Lax1971, Lax1972}. In 1964, Mark Wilkins published the artificial viscosity in the form that is most widely used\cite{Wilkins1964}. In the same volume, William Schulz described a tensor viscosity for use with a two-dimensional Lagrangian code\cite{Schulz1964}. The stability analysis techniques and convergence theory of partial differential equations were established by von Neumann, Richtmyer, and Lax\cite{30, vonNeumannRichtmyer1947, LaxRichtmyer1956}. See Maattsson and Rider for a review of the development of artificial viscosity\cite{MattssonRider2015}.

The creation of shock capturing schemes facilitated the development of two-dimensional Lagrangian shock hydrodynamics codes\cite{50, Kolsky2D, Hermann, Schulz1964, MaenchenSack1964, Wilkins1964}, and eventually three-dimensional codes\cite{BurtonFLAG2007, ALE3D2017}. Modern staggered-grid Lagrangian and artificial viscosity methods have evolved from the pioneering work of the 1940's\cite{BurtonSGH, Caramana1998a, Caramana1998b, Caramana1998c, Caramana2000, CampbellShashkov2001, Bauer2006, Barlow2008}. The first cell-centered Lagrangian hydrodynamic method was proposed by Sergei Godunov\cite{Godunov1, Godunov2}.  Research at LANL on cell-centered Lagrangian hydrodynamic methods in the early 1980's resulted in the CAVEAT code\cite{CAVEAT, CAVEATGT}, which was followed by a myriad of work on cell-centered schemes in the 2000's\cite{Despres, Maire2007, Maire2008, Maire2009, Carre2009, Loubere2010a, Barlow2011, Burton2013, BurtonCGR, ChiravalleCCH, MorganVeloFilter2017}. The cell-centered work has led to notable improvements in the staggered-grid methods\cite{Loubere2010b, Maire3, Barlow2013, Loubere2013, MorganMARS, ChiravalleMARS}.

This history shows the vital contributions made to the hydrodynamic methods theory by von Neumann, Peierls, Skyrme, and Richtmyer. It certainly justifies Hans Bethe's statement that the ``collaboration of the British Mission absolutely was essential''\cite{51} and Norris Bradbury's statement that ``the British Mission supplied the major portion of experience in the field of theoretical hydrodynamics... It might also be pointed out that the United States was largely lacking in personnel experienced in this field of classical physics''\cite{51}. 
This paper sought to correct historical omissions and to communicate the complicated story on the origins of Lagrangian hydrodynamic methods for simulating shock dynamics.

\section{Acknowledgements}
We are thankful for the helpful advice and input from Bill Rider, Don Burton, Len Margolin, and Misha Shashkov.  We also gratefully acknowledge the support from the Advanced Scientific Computing (ASC) program at Los Alamos National Laboratory. The Los Alamos unlimited release number is LA-UR-21-20144. Los Alamos National Laboratory is operated by Triad National Security, LLC for the U.S. Department of Energy NNSA under Contract No. 89233218CNA000001.

\bibliographystyle{ieeetr} 
\bibliography{refs.bib}

\begin{thebibliography}{100}

\bibitem{1}
T.~O. of~Scientific~Research and D.~Collection, ``Library of congress, science
  reference reports, technical reports and standards.''
\newblock https://www.loc.gov/rr/scitech/trs/trsosrd.html.

\bibitem{2}
H.~Bethe, ``The theory of shock waves for an arbitrary equation of state,''
  Tech. Rep. OSRD No. 545, Serial No. 237, Division 8, National Defense
  Research Committee of the Office of Scientific Research and Development, May
  1942.
\newblock Report.

\bibitem{3}
J.~Kirkwood and H.~Bethe, ``The pressure wave produced by an underwater
  explosion, part {I}.,'' May 1942.
\newblock Report.

\bibitem{4}
J.~Kirkwood, H.~Bethe, and E.~Montroll, ``The pressure wave produced by an
  underwater explosion, part {I}{I}.,'' May 1942.
\newblock Report.

\bibitem{5}
J.~von Neumann Tech. Rep. AM-9, NDRC, Div. B, June 1941.
\newblock Report.

\bibitem{6}
J.~von Neumann, ``Theory of detonation waves,'' Tech. Rep. OSRD No. 549,
  Institute for Advanced Study, April 1942.
\newblock Report.

\bibitem{7}
J.~von Neumann, ``Theory of shock waves,'' Tech. Rep. OSRD No. 1140, Institute
  for Advanced Study, January 1943.
\newblock Report.

\bibitem{8}
J.~von Neumann, ``Oblique reflection of shock waves,'' tech. rep., Naval
  Ordnance, Explosive Research Report No. 12, October 1943.

\bibitem{9}
J.~B.~Wilson, ``Effects of impact and explosion,'' 1946.

\bibitem{10}
W.~Penney and H.~Pike, ``Shock waves and the propagation of finite pulses in
  fluids,'' {\em Reports on Progress in Physics}, vol.~13, no.~1, p.~46, 1950.

\bibitem{11}
R.~Peierls, {\em Bird of passage: recollections of a physicist}, vol.~55.
\newblock Princeton University Press, 2014.

\bibitem{12}
G.~Taylor, ``British report {R}{C}-210,'' June 1941.

\bibitem{13}
G.~Taylor, ``Notes on the dynamics of shock waves from bare explosive
  charges,'' 1941.

\bibitem{14}
G.~Taylor, ``Detonation waves,'' 1941.

\bibitem{15}
R.~Serber, ``Exponential shock and rarefaction waves,'' Tech. Rep. LA-14, Los
  Alamos Scientific Laboratory, July 1943.
\newblock Report.

\bibitem{ArcherComputers}
B.~Archer, ``The los alamos computing facility during the manhattan project,''
  {\em submitted to ANS Nuclear Technology special issue}, 2021.

\bibitem{18}
D.~Hawkins, ``Manhattan district history project {Y} the los alamos project,
  {V}ol. {I} inception until august 1945,'' Tech. Rep. LAMS-2532 (Vol. I), Los
  Alamos Scientific Laboratory, December 1961.
\newblock written 1946 and 1947.

\bibitem{22}
J.~von Neumann, ``Proposal and analysis of a new numerical method for the
  treatment of hydrodynamical shock problems,'' March 1944.
\newblock Applied Mathematics Panel National Defense Research Committee Report.

\bibitem{31}
T.~Skyrme, ``Treatment of discontinuities in {L}agrangian integration of
  symmetrical hydrodynamic problems,'' Tech. Rep. LAMS-562, Los Alamos
  Scientific Laboratory, October 1948.
\newblock Report.

\bibitem{37}
R.~Peierls, ``{Theory on von Neumann’s method of treating shocks},'' Tech.
  Rep. LA-332, Los Alamos Scientific Laboratory, 1945.
\newblock Report.

\bibitem{23}
L.~Hoddeson, P.~Henriksen, R.~Meade, and C.~Westfall, {\em Critical assembly: a
  technical history of Los Alamos during the Oppenheimer years, 1943-1945}.
\newblock Cambridge University Press, 1993.

\bibitem{24}
A.~K. Smith and C.~Weiner, {\em Robert Oppenheimer: Letters and Recollections}.
\newblock Harvard University Press, 1980.

\bibitem{CFL1928}
R.~Courant, K.~Friedrichs, and H.~Lewy, ``{\"U}ber die partiellen
  differenzengleichungen der mathematischen physik,'' {\em Mathematische
  annalen}, vol.~100, no.~1, pp.~32--74, 1928.

\bibitem{30}
J.~von Neumann and R.~Richtmyer, ``A method for the numerical calculation of
  hydrodynamic shocks,'' {\em Journal of applied physics}, vol.~21, no.~3,
  pp.~232--237, 1950.

\bibitem{25}
B.~Archer, ``Peierls' letters in march 1944,'' Tech. Rep. LA-UR-20-28408, Los
  Alamos National Laboratory, October 2020.
\newblock Report.

\bibitem{35}
E.~Nelson, T.~Skyrme, R.~Peierls, S.~Goldberg, J.~Kemeny, D.~Flanders, and
  P.~Whitman, ``Numerical methods, part {I}{I}, {C}hapters 6 thru 8,'' Tech.
  Rep. LA-1058, Los Alamos Scientific Laboratory, June 1949.
\newblock Report.

\bibitem{RichtmyerHydro}
R.~D. Richtmyer, ``Proposed numerical method for calculation of shocks,'' Tech.
  Rep. LA-671, Los Alamos Scientific Laboratory, March 1948.
\newblock Report.

\bibitem{Emmons1}
H.~Emmons, ``The numerical solution of compressible fluid flow problems,''
  Tech. Rep. NACA Tech. Note No. 932, 1944.

\bibitem{Emmons2}
H.~Emmons, ``Flow of a compressible fluid past a symmetrical airfoil in a wind
  tunnel and in free air,'' Tech. Rep. NACA Tech. Note No. 1746, 1948.

\bibitem{DIGIACINTO198961}
M.~D. Giacinto and M.~Valorani, ``Shock detection and discontinuity tracking
  for unsteady flows,'' {\em Computers and Fluids}, vol.~17, no.~1, pp.~61--84,
  1989.

\bibitem{GLIMM1984}
J.~Glimm, C.~Klingenberg, O.~McBryan, B.~Plohr, D.~Sharp, and S.~Yaniv, ``Front
  tracking and two dimensional {R}iemann problems: A conference report,'' Tech.
  Rep. LA-UR-84-3209, Los Alamos National Laboratory, 1984.

\bibitem{SHARP1993}
D.~Sharp, J.~Grove, Y.~Yang, B.~Boston, R.~Holmes, and Q.~Zhang, ``The
  application of front tracking to the simulation of shock refractions and
  shock accelerated interface mixing,'' in {\em Proceedings of the 4th
  International Workshop on the Physics of Compressible Turbulent Mixing,
  Cambridge University, Cambridge England. State Univ. of New York at Stony
  Brook}, 1993.

\bibitem{GLIMM2013}
J.~Glimm, D.~H. Sharp, T.~Kaman, and H.~Lim, ``New directions for
  rayleigh--taylor mixing,'' {\em Philosophical Transactions of the Royal
  Society A: Mathematical, Physical and Engineering Sciences}, vol.~371,
  no.~2003, p.~20120183, 2013.

\bibitem{ONOFRI2017}
M.~Onofri and R.~Paciorri, {\em Shock Fitting: Classical Techniques, Recent
  Developments, and Memoirs of Gino Moretti}.
\newblock Springer, 2017.

\bibitem{mDG-ICE-AndrewC}
A.~T. Corrigan, A.~Kercher, and D.~A. Kessler, ``The moving discontinuous
  galerkin method with interface condition enforcement for unsteady
  three-dimensional flows,'' {\em AIAA Scitech 2019 Forum}, 2019.

\bibitem{mDG-ICE-AndrewK}
A.~D. Kercher, A.~Corrigan, and D.~A. Kessler, ``The moving discontinuous
  galerkin finite element method with interface condition enforcement for
  compressible viscous flows,'' {\em International Journal for Numerical
  Methods in Fluids}, 2020.

\bibitem{26}
L.~Badash, Hirschfelder, and H.~Broida, {\em {Reminiscences of Los Alamos,
  1943--1945}}.
\newblock D. Reidel Publishing Company, Dordrecht, Holland, 1980.

\bibitem{28}
N.~Metropolis and E.~C. Nelson, ``{Early computing at Los Alamos},'' {\em
  Annals of the History of Computing}, vol.~4, no.~4, pp.~348--357, 1982.

\bibitem{49}
R.~Richtmyer, ``Proposed numerical method for calculation of shocks, {I}{I},''
  Tech. Rep. LA-699, Los Alamos Scientific Laboratory, 1948.

\bibitem{39}
A.~T. S.~Fembach, ``Computers and their role in the physical sciences,'' 1970.

\bibitem{Sedov}
L.~Sedov, ``Similarity and dimensional methods in mechanics,'' {\em Academic
  Press}, 1959.

\bibitem{MorganSedov}
C.~Pederson, B.~Brown, and N.~Morgan, ``The {S}edov blast wave as a radial
  piston verification test,'' {\em Journal of Verification, Validation and
  Uncertainty Quantification}, vol.~1, no.~3, 2016.

\bibitem{50}
H.~Kolsky, I.~Cherry, R.~Clark, P.~Harper, R.~Oeder, and R.~Stark,
  ``{Preliminary description of the method used in the Model II C.P.C.
  calculations of T-5's Standard Hydrodynamics Problems},'' Tech. Rep.
  LAMS-1572, Los Alamos Scientific Laboratory, March 1953.
\newblock Report.

\bibitem{VilarDGlagcaf2012}
F.~Vilar, ``{Cell-centered discontinuous Galerkin discretization for
  two-dimensional Lagrangian hydrodynamics},'' {\em Computers \& fluids},
  vol.~64, pp.~64--73, 2012.

\bibitem{VilarDG}
F.~Vilar, P.-H. Maire, and R.~Abgrall, ``{A discontinuous Galerkin
  discretization for solving the two-dimensional gas dynamics equations written
  under total Lagrangian formulation on general unstructured grids},'' {\em
  Journal of Computational Physics}, vol.~276, pp.~188--234, 2014.

\bibitem{LiuDGlagcaf2017}
X.~Liu, N.~R. Morgan, and D.~E. Burton, ``{A Lagrangian discontinuous Galerkin
  hydrodynamic method},'' {\em Computers \& Fluids}, vol.~163, pp.~68--85,
  2018.

\bibitem{LiuDGlagSMS2019}
X.~Liu, N.~R. Morgan, and D.~E. Burton, ``{A high-order Lagrangian
  discontinuous Galerkin hydrodynamic method for quadratic cells using a
  subcell mesh stabilization scheme},'' {\em Journal of Computational Physics},
  vol.~386, pp.~110--157, 2019.

\bibitem{LiuDGlagRZ2018}
X.~Liu, N.~R. Morgan, and D.~E. Burton, ``{Lagrangian discontinuous Galerkin
  hydrodynamic methods in axisymmetric coordinates},'' {\em Journal of
  Computational Physics}, vol.~373, pp.~253--283, 2018.

\bibitem{LiebermanStrength1D2017}
E.~J. Lieberman, N.~R. Morgan, D.~J. Luscher, and D.~E. Burton, ``{A
  higher-order Lagrangian discontinuous Galerkin hydrodynamic method for
  elastic-plastic flows},'' {\em Computers $\&$ Fluids}, vol.~78, pp.~318--334,
  2019.

\bibitem{LIEBERMAN2019467}
E.~J. Lieberman, X.~Liu, N.~R. Morgan, D.~J. Luscher, and D.~E. Burton, ``{A
  higher-order Lagrangian discontinuous Galerkin hydrodynamic method for solid
  dynamics},'' {\em Computer Methods in Applied Mechanics and Engineering},
  vol.~353, pp.~467--490, 2019.

\bibitem{DGHEburn}
E.~J. Lieberman, X.~Liu, N.~R. Morgan, and D.~E. Burton, ``{A multiphase
  Lagrangian discontinuous Galerkin hydrodynamic method for high-explosive
  detonation physics},'' {\em Applications in Engineering Science}, vol.~4,
  p.~100022, 2020.

\bibitem{BLAST}
V.~Dobrev, T.~Ellis, T.~V. Kolev, and R.~Rieben, ``{Curvilinear finite elements
  for Lagrangian hydrodynamics},'' {\em International Journal for Numerical
  Methods in Fluids}, vol.~65, no.~11-12, pp.~1295--1310, 2011.

\bibitem{SvetlanaRD}
R.~Abgrall, K.~Lipnikov, N.~Morgan, and S.~Tokareva, ``Multidimensional
  staggered grid residual distribution scheme for {L}agrangian hydrodynamics,''
  {\em SIAM Journal on Scientific Computing}, vol.~42, no.~1, pp.~A343--A370,
  2020.

\bibitem{ANN-ShockDetector}
N.~R. Morgan, S.~Tokareva, X.~Liu, and A.~Morgan, ``A machine learning approach
  for detecting shocks with high-order hydrodynamic methods,'' in {\em 2020
  AIAA Aerospace Sciences Meeting}, p.~2024, 2020.

\bibitem{Kolsky2D}
H.~Kolsky, ``A method for the numerical solution of transient hydrodynamic
  shock problems in two space dimensions,'' Tech. Rep. LA-1867, Los Alamos
  Scientific Laboratory, 1955.
\newblock Report.

\bibitem{BensonReview1992}
D.~J. Benson, ``{Computational methods in Lagrangian and Eulerian
  hydrocodes},'' {\em Computer methods in Applied mechanics and Engineering},
  vol.~99, no.~2-3, pp.~235--394, 1992.

\bibitem{WilkinsSGH}
M.~Wilkins, ``Use of artificial viscosity in multidimensional shock wave
  problems,'' {\em J. Comput. Phys}, vol.~36, no.~3, pp.~281--303, 1980.

\bibitem{Zukas2004}
J.~Zukas, {\em Introduction to hydrocodes}.
\newblock Elsevier, 2004.

\bibitem{BurtonFLAG2007}
D.~E. Burton, ``{Lagrangian hydrodynamics in the FLAG code},'' {\em Los Alamos
  National Laboratory, Los Alamos, NM, Technical Report No. LA-UR-07-7547},
  2007.

\bibitem{ALE3D2017}
C.~R. Noble, A.~T. Anderson, N.~R. Barton, J.~A. Bramwell, A.~Capps, M.~H.
  Chang, J.~J. Chou, D.~M. Dawson, E.~R. Diana, T.~A. Dunn, {\em et~al.},
  ``{ALE3D: An arbitrary Lagrangian-Eulerian multi-physics code},'' tech. rep.,
  Lawrence Livermore National Laboratory, 2017.

\bibitem{BurtonSGH}
D.~Burton, ``Multidimensional discretization of conservation laws for
  unstructured polyhedral grids,'' tech. rep., LLNL, UCRL-JC-118306, 1994.

\bibitem{Caramana}
E.~Caramana, D.~Burton, M.~J. Shashkov, and P.~Whalen, ``The construction of
  compatible hydrodynamic algorithms utilizing conservation of total energy,''
  {\em Journal of Applied Physics}, vol.~146, no.~1, pp.~227--262, 1998.

\bibitem{Landshoff1955}
R.~Landshoff, ``A numerical method for treating fluid flow in the presence of
  shocks,'' tech. rep., Los Alamos Scientific Laboratory, 1955.

\bibitem{MorganSGHdisp}
N.~R. Morgan, ``{A dissipation model for staggered grid Lagrangian
  hydrodynamics},'' {\em Computers \& Fluids}, vol.~83, pp.~48--57, 2013.

\bibitem{RichtmyerMorton1967}
R.~D. {Richtmyer} and K.~W. {Morton}, {\em {Difference methods for
  initial-value problems, Second Edition}}.
\newblock John Wiley \& Sons, Inc., New York, New York, 1967.

\bibitem{Lax1954}
P.~D. Lax, ``Weak solutions of nonlinear hyperbolic equations and their
  numerical computation,'' {\em Communications on pure and applied
  mathematics}, vol.~7, no.~1, pp.~159--193, 1954.

\bibitem{LaxWendroff1958}
P.~Lax, ``Systems of conservation laws,'' tech. rep., Los Alamos Scientific
  Laboratory, 1958.

\bibitem{LaxWendroff1960}
B.~W. P.~Lax, ``Systems of conservation laws,'' {\em Communications on Pure and
  Applied Mathematics}, vol.~13, 1960.

\bibitem{Lax1971}
P.~Lax, ``Shock waves and entropy,'' in {\em Contributions to nonlinear
  functional analysis} (H.~Zarantonello, ed.), Academic Press, 1971.

\bibitem{Lax1972}
P.~Lax, {\em Hyperbolic systems of conservation laws and the mathematical
  theory of shock waves}.
\newblock SIAM, 1973.

\bibitem{Wilkins1964}
M.~Wilkins, ``Calculation of elastic-plastic flow,'' in {\em Methods in
  Computational Physics, Advances in Research and Applications, Volume 3,
  Fundamental Methods in Hydrodynamics}, Academic Press, 1964.

\bibitem{Schulz1964}
W.~Schulz, ``{Two-Dimensional Lagrangian Hydrodynamic Difference Equations},''
  in {\em Methods in Computational Physics, Advances in Research and
  Applications, Volume 3, Fundamental Methods in Hydrodynamics}, Academic
  Press, 1964.

\bibitem{vonNeumannRichtmyer1947}
A.~Carr, ``On the numerical solution of partial differential equations of
  parabolic type,'' tech. rep., Los Alamos Scientific Laboratory, report,
  LA-657, 1947.

\bibitem{LaxRichtmyer1956}
P.~D. Lax and R.~D. Richtmyer, ``Survey of the stability of linear finite
  difference equations,'' {\em Communications on pure and applied mathematics},
  vol.~9, no.~2, pp.~267--293, 1956.

\bibitem{MattssonRider2015}
A.~E. Mattsson and W.~J. Rider, ``Artificial viscosity: back to the basics,''
  {\em International Journal for Numerical Methods in Fluids}, vol.~77, no.~7,
  pp.~400--417, 2015.

\bibitem{Hermann}
W.~Herrmann, ``Comparison of finite difference expressions used in {L}agrangian
  fluid flow calculations,'' Tech. Rep. WL-TR-64-104, Air Force Weapons
  Laboratory, 1964.
\newblock Report.

\bibitem{MaenchenSack1964}
G.~Maenchen and S.~Sacks, ``The tensor code, in methods of computational
  physics, 3, 181-210. b. alder, s. fernback and m. rotenberg editors,'' 1964.

\bibitem{Caramana1998a}
E.~Caramana and M.~Shashkov, ``{Elimination of artificial grid distortion and
  hourglass-type motions by means of Lagrangian subzonal masses and
  pressures},'' {\em Journal of Computational Physics}, vol.~142, no.~2,
  pp.~521--561, 1998.

\bibitem{Caramana1998b}
E.~J. Caramana, M.~J. Shashkov, and P.~P. Whalen, ``Formulations of artificial
  viscosity for multi-dimensional shock wave computations,'' {\em Journal of
  Computational Physics}, vol.~144, no.~1, pp.~70--97, 1998.

\bibitem{Caramana1998c}
E.~Caramana, D.~Burton, M.~J. Shashkov, and P.~Whalen, ``The construction of
  compatible hydrodynamics algorithms utilizing conservation of total energy,''
  {\em Journal of Computational Physics}, vol.~146, no.~1, pp.~227--262, 1998.

\bibitem{Caramana2000}
E.~Caramana, C.~Rousculp, and D.~Burton, ``{A compatible, energy and symmetry
  preserving Lagrangian hydrodynamics algorithm in three-dimensional Cartesian
  geometry},'' {\em Journal of Computational Physics}, vol.~157, no.~1,
  pp.~89--119, 2000.

\bibitem{CampbellShashkov2001}
J.~Campbell and M.~J. Shashkov, ``A tensor artificial viscosity using a mimetic
  finite difference algorithm,'' {\em Journal of Computational Physics},
  vol.~172, no.~2, pp.~739--765, 2001.

\bibitem{Bauer2006}
A.~L. Bauer, D.~E. Burton, E.~Caramana, R.~Loub{\`e}re, M.~J. Shashkov, and
  P.~Whalen, ``{The internal consistency, stability, and accuracy of the
  discrete, compatible formulation of Lagrangian hydrodynamics},'' {\em Journal
  of Computational Physics}, vol.~218, no.~2, pp.~572--593, 2006.

\bibitem{Barlow2008}
A.~J. Barlow, ``A compatible finite element multi-material ale hydrodynamics
  algorithm,'' {\em International journal for numerical methods in fluids},
  vol.~56, no.~8, pp.~953--964, 2008.

\bibitem{Godunov1}
S.~Godounov, ``R{\'e}solution num{\'e}rique des problemes multidimensionnels de
  la dynamique des gaz,'' 1979.

\bibitem{Godunov2}
S.~K. Godunov, ``Reminiscences about difference schemes,'' {\em Journal of
  Computational Physics}, vol.~153, no.~1, pp.~6--25, 1999.

\bibitem{CAVEAT}
F.~L. Addessio, J.~R. Baumgardner, J.~K. Dukowicz, N.~L. Johnson, B.~A.
  Kashiwa, R.~M. Rauenzahn, and C.~Zemach, ``{CAVEAT: A computer code for fluid
  dynamics problems with large distortion and internal slip. Revision 1},''
  Tech. Rep. LA-10613-MS-REV.1, Los Alamos National Laboratory, 1990.

\bibitem{CAVEATGT}
F.~L. Addessio, M.~Cline, and J.~K. Dukowicz, ``A general topology, godunov
  method,'' {\em Computer Physics Communications}, vol.~48, no.~1, pp.~65--73,
  1988.

\bibitem{Despres}
B.~Despr{\'e}s and C.~Mazeran, ``Lagrangian gas dynamics in two dimensions and
  lagrangian systems,'' {\em Archive for Rational Mechanics and Analysis},
  vol.~178, no.~3, pp.~327--372, 2005.

\bibitem{Maire2007}
P.-H. Maire, R.~Abgrall, J.~Breil, and J.~Ovadia, ``{A cell-centered Lagrangian
  scheme for two-dimensional compressible flow problems},'' {\em SIAM Journal
  on Scientific Computing}, vol.~29, no.~4, pp.~1781--1824, 2007.

\bibitem{Maire2008}
P.-H. Maire and J.~Breil, ``{A second-order cell-centered Lagrangian scheme for
  two-dimensional compressible flow problems},'' {\em International journal for
  numerical methods in fluids}, vol.~56, no.~8, pp.~1417--1423, 2008.

\bibitem{Maire2009}
P.-H. Maire, ``{A high-order cell-centered Lagrangian scheme for
  two-dimensional compressible fluid flows on unstructured meshes},'' {\em
  Journal of Computational Physics}, vol.~228, no.~7, pp.~2391--2425, 2009.

\bibitem{Carre2009}
G.~Carr{\'e}, S.~Del~Pino, B.~Despr{\'e}s, and E.~Labourasse, ``{A
  cell-centered Lagrangian hydrodynamics scheme on general unstructured meshes
  in arbitrary dimension},'' {\em Journal of Computational Physics}, vol.~228,
  no.~14, pp.~5160--5183, 2009.

\bibitem{Loubere2010a}
R.~Loub{\`e}re, P.-H. Maire, M.~Shashkov, J.~Breil, and S.~Galera, ``{ReALE: a
  reconnection-based arbitrary-Lagrangian--Eulerian method},'' {\em Journal of
  Computational Physics}, vol.~229, no.~12, pp.~4724--4761, 2010.

\bibitem{Barlow2011}
A.~Barlow and P.~Roe, ``{A cell centred Lagrangian Godunov scheme for shock
  hydrodynamics},'' {\em Computers \& fluids}, vol.~46, no.~1, pp.~133--136,
  2011.

\bibitem{Burton2013}
D.~Burton, T.~Carney, N.~Morgan, S.~Sambasivan, and M.~Shashkov, ``{A
  cell-centered Lagrangian Godunov-like method for solid dynamics},'' {\em
  Computers \& Fluids}, vol.~83, pp.~33--47, 2013.

\bibitem{BurtonCGR}
D.~E. Burton, N.~R. Morgan, T.~C. Carney, and M.~A. Kenamond, ``{Reduction of
  dissipation in Lagrange cell-centered hydrodynamics (CCH) through corner
  gradient reconstruction (CGR)},'' {\em Journal of Computational Physics},
  vol.~299, pp.~229--280, 2015.

\bibitem{ChiravalleCCH}
V.~P. Chiravalle and N.~R. Morgan, ``{A 3D Lagrangian cell-centered
  hydrodynamic method with higher-order reconstructions for gas and solid
  dynamics},'' {\em Computers \& Fluids}, vol.~83, pp.~642--663, 2018.

\bibitem{MorganVeloFilter2017}
N.~R. Morgan, X.~Liu, and D.~E. Burton, ``{Reducing spurious mesh motion in
  Lagrangian finite volume and discontinuous Galerkin hydrodynamic methods},''
  {\em Journal of Computational Physics}, vol.~372, pp.~35--61, 2018.

\bibitem{Loubere2010b}
R.~Loub{\`e}re, P.-H. Maire, and P.~V{\'a}chal, ``{A second-order compatible
  staggered Lagrangian hydrodynamics scheme using a cell-centered
  multidimensional approximate Riemann solver},'' {\em Procedia Computer
  Science}, vol.~1, no.~1, pp.~1931--1939, 2010.

\bibitem{Maire3}
P.-H. Maire, R.~Loubere, and P.~V{\'a}chal, ``{Staggered Lagrangian
  discretization based on cell-centered Riemann solver and associated
  hydrodynamics scheme},'' {\em Communications in Computational Physics},
  vol.~10, no.~4, pp.~940--978, 2011.

\bibitem{Barlow2013}
A.~Barlow, ``{A high order cell centred dual grid Lagrangian Godunov scheme},''
  {\em Computers \& Fluids}, vol.~83, pp.~15--24, 2013.

\bibitem{Loubere2013}
R.~Loub{\`e}re, ``Contribution au domaine des m{\`e}thodes num{\`e}riques
  {L}agrangiennes et arbitrary-{L}agrangian--{E}ulerian,'' {\em Habilitation
  {\`a} Diriger Des Recherche, University of Toulouse}, 2013.

\bibitem{MorganMARS}
N.~R. Morgan, K.~N. Lipnikov, D.~E. Burton, and M.~A. Kenamond, ``{A Lagrangian
  staggered grid Godunov-like approach for hydrodynamics},'' {\em Journal of
  Computational Physics}, vol.~259, pp.~568--597, 2014.

\bibitem{ChiravalleMARS}
V.~Chiravalle and N.~Morgan, ``{A 3D finite element ALE method using an
  approximate Riemann solution},'' {\em International Journal for Numerical
  Methods in Fluids}, vol.~83, no.~8, pp.~642--663, 2016.

\bibitem{51}
A.~Carr, ``Documents pertaining to the {B}ritish mission,'' tech. rep., LANL,
  report, LA-UR-09-0554, 2009.

\end{thebibliography}

\appendix

\section{Original Letters}
\label{app:letters}
This appendix provides the original letters sent by Rudolf Peierls to J.R. Oppenheimer and J. von Neumann in March 1944. These letters were declassified and approved for unlimited release, LA-UR-20-28408. Peierls made his first visit to Los Alamos on February 8, 1944, and subsequently moved to Los Alamos in June 1944 to lead the implosion theory group. The letters are the result of the discussions during that first visit to Los Alamos. These letters were found in the archives of the Manhattan Project held at the Los Alamos National Laboratory National Security Research Center.

The first four letters are to J.R. Oppenheimer.
\begin{itemize}
    \item March 9, 1944 discusses neutronics and hydrodynamics.
    \item March 15, 1944 mainly discusses hydrodynamics, giving the shock wave equations worked out by T.H.R. Skyrme. Skyrme joined Los Alamos on July 31, 1944 and was a key member of the hydrodynamics modeling effort.
    \item March 21, 1944 discuses neutronics by Davison.
    \item March 24, 1944 discuses neutronics by Wilson.
\end{itemize}

The last letter is to J. von Neumann on March 28, 1944 discussing Peierls’ method for handling shock waves. The last paragraph of this letter is remarkable because Peierls suggests to von Neumann to use artificial viscosity to smear out shocks in hydrodynamics calculations.



\section*{Supplementary material}

# THE BRITISH SUPPLY COUNCIL IN NORTH AMERICA

TELEPHONE
HAnover 2-2460

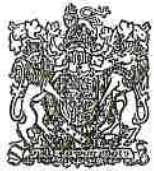

P. O. Box 51
WALL STREET STATION
NEW YORK 5, N. Y.

9th March, 1944.

Dr. J. R. Oppenheimer,
P. O. Box 1663,
Santa Fe, New Mexico.

Dear Oppenheimer,

I can now give you answers to some of the questions that I put to Wilson as a result of our recent discussions.

(a) <u>Perturbation Theory for Small Deviations from Spherical Shape.</u>

Wilson has looked into this and is writing up the theory. I understand from Mr. Skyrme, who has just arrived here from Birmingham, that Wilson considers this a very easy problem and, in fact, was in doubt whether there was not some more subtle point in our minds when we made the request, since the method he suggests to solve the problem is fairly obvious. I hope to get an account of this method from him shortly and shall, of course, then forward it to you at once. We can then see whether this takes care of the problem Bethe was anxious to get solved.

(b) <u>Numerical Method for the Shock Wave Problem.</u>

The present state of affairs is as follows: Most of the time lately has been spent on attempting to solve the equations for small deviations from the Neuman case to a high order of approximation, but it turned out that the equations for all but the first order become unmanageable. We believe, however, that a solution up to first order should give a sufficiently good basis for starting numerical integration and this had just started again when Mr. Skyrme left. It was, however, then too early to say what further difficulties might be encountered. In the meantime work has been done on finding an analytical form for the equation of state of air in the range concerned. This work was almost completed when Skyrme left but as he was working on this himself his departure may have caused some delay. I assume that, regardless of the fate of the numerical integration, this work on the equation of state should be pushed ahead as it would also be of use for Bethe's approximate method of solution. It is difficult to judge the time scale for this integration process as it may be fairly rapid if there are no snags. Our interpretation is that it should easily be completed in a time of the order of six months if everything runs smoothly, but it is quite possible that further difficulties may cause an appreciable delay.

Dr. J. R. Oppenheimer  -2-  9th March, 1944.

(c) New Form of the Equations for Numerical Work.

Mr. Skyrme has developed an alternative form of the equations in which the variables are $r$ and $r/Y$ where $Y$ is the shock wave radius. This leads to fairly simple equations and to boundary conditions which are easier than the usual ones. Skyrme will write down the equations which I will then send on to you. I don't believe that this method will be of particular interest in connection with your integration problem but it might give some improvement for the latter stages after part of the material has started moving outwards and when there is a definite shock wave.

I hope you have in the meantime received the latest reports from Wilson to which I referred in a previous letter. These were transmitted to your Washington Office on February 9th. You may know that Chadwick has a full list of British Reports handed over, together with the dates when they were passed on. He may be able to give you information that would help to chase up any particular reports.

Yours sincerely,

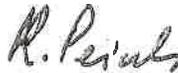

R. Peierls

CLASSIFICATION CANCELLED
PER DOC REVIEW JAN. 1973

~~U. S. SECRET~~
~~BRITISH MOST SECRET~~

RP:jeg.                                         c.c. Dr. Webster

# THE BRITISH SUPPLY COUNCIL IN NORTH AMERICA

TELEPHONE
HAnover 2-2460

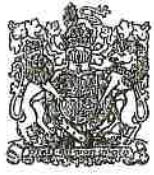

P. O. Box 51
WALL STREET STATION
NEW YORK 5, N. Y.

CLASSIFICATION CANCELLED
PER DOC REVIEW JAN. 197-

15th March, 1944.

Dr. J. R. Oppenheimer,
P. O. Box 1663,
Santa Fe, New Mexico.

U. S. SECRET
BRITISH MOST SECRET

Dear Oppenheimer,

I have received an advance copy of the February Progress Report of the Birmingham Group. This, of course, will eventually reach you through official channels but as the final copy of the January Report does not even seem to have reached this country, these are evidently very slow and I want to mention to you 2 items that may interest you.

On the shock wave problem, numerical integration has been resumed and led at first to disappointing results as the accuracy obtained seemed poor. They have stopped to investigate the reason for this and will report to us any further results. In the meantime, work on the equation of state for air is continuing. They have worked out the general theory of a hemisphere completely surrounded by an infinite scattering container and the results are now being evaluated numerically. I would welcome some indication on whether you think this problem at all of interest and whether it should be pressed with any urgency.

Dirac reports that his calculations on efficiency have now been generalized to include the scattering in the container and extensive numerical work is now starting on this. This report is dated about two weeks after I cabled to Dirac, warning him about the error which Bethe has found in his method and I am not quite clear whether in the numerical work now beginning this has been corrected for. In any case I think it would be very helpful indeed if we could before long have the notes which Bethe promised to write up on this point.

I mentioned in my last letter the new form of the shock wave equations which was suggested by Skyrme. We have looked into the question of their suitability for numerical work and come to the conclusion that for purposes of numerical equations they are completely equivalent to the old ones. It is, however, of some interest that, in the variables Skyrme uses, the equations are rather simpler than I would have expected and I enclose a note on the form of the equations for what it is worth.

Yours sincerely,

R. Peierls

enc.

cc - Theoretical phys.

**VERIFIED UNCLASSIFIED**

P. M. Lang, OS-6

MAY 1 1986

# An alternative form of the equation of a spherical shock wave

In the usual notation

$\underline{R}$ is position of particle initially at $\underline{r}$ at time $\underline{t}$.

$p, \rho$ are pressure and density

$\rho_0$ is initial density outside disturbance

The equations are then

$$\frac{\partial^2 R}{\partial t^2} = - \frac{R^2}{r^2} \frac{\partial \Pi}{\partial r} \qquad (1)$$

$$g = \frac{R^2}{r^2} \frac{\partial R}{\partial r} \qquad (2)$$

where
$$\Pi = p/\rho_0 \quad ; \quad g = \rho_0/\rho \qquad (3)$$

<u>Boundary conditions</u>; when $r = 0$; $R = \dot{R} = 0$ \qquad (4)

when $\quad R = r = Y(t)$ \qquad (5)

$$\Pi = \dot{Y} u \qquad (6)$$

$$\dot{Y} g = \dot{Y} - u \qquad (7)$$

$$2E/\Pi g = 1/g - 1 \qquad (8)$$

and $\quad u = \frac{\partial R}{\partial t} \qquad (9)$

Take as variable instead of $t$, $z = r/Y$

In terms of $z$ & $r$

$$\frac{\partial}{\partial t} = \frac{\partial z}{\partial t} \frac{\partial}{\partial z} = - \frac{\dot{Y}}{Y} z \frac{\partial}{\partial z} \qquad (10)$$

$$r \frac{\partial}{\partial r} = z \frac{\partial}{\partial z} + r \frac{\partial}{\partial r} \qquad (11)$$

Write
$$\left. \begin{array}{l} R = Y\, F(z, r) \\ \Pi = \dot{Y}^2\, P(z, r) \\ g = g(z, r) \\ u = \dot{Y}\, H(z, r) \end{array} \right\} \qquad (12)$$

Then $u = \dfrac{\partial R}{\partial t}$ becomes

$$\dot{y} H = \dot{y} \bar{F} + y\left(-z \dfrac{\dot{y}}{y} \dfrac{\partial \bar{F}}{\partial z}\right)$$

or
$$H = \bar{F} - z \dfrac{\partial \bar{F}}{\partial z} \qquad (13)$$

and $\dot{u} = -\dfrac{R^2}{r^2} \dfrac{\partial \Pi}{\partial r}$ becomes

$$\ddot{y} H + \dot{y}\left(-z \dfrac{\dot{y}}{y} \dfrac{\partial H}{\partial z}\right) = -\dfrac{y^2 \bar{F}^2}{r^2} \cdot \dfrac{1}{r} \dot{y}^2 \left(z \dfrac{\partial P}{\partial z} + r \dfrac{\partial P}{\partial r}\right) \qquad (14)$$

or
$$\left(z \dfrac{\partial}{\partial z} + r \dfrac{\partial}{\partial r}\right) P = z^3/\bar{F}^2 \left(z \dfrac{\partial H}{\partial z} - \dfrac{y \cdot \ddot{y}}{\dot{y}^2} H\right)$$

and $g = \dfrac{R^2}{r^2} \dfrac{\partial R}{\partial r}$ becomes

$$g = \dfrac{y^2 \bar{F}^2}{r^2} \cdot \dfrac{y}{r} \left(z \dfrac{\partial \bar{F}}{\partial z} + r \dfrac{\partial \bar{F}}{\partial r}\right)$$

or
$$\left(z \dfrac{\partial}{\partial z} + r \dfrac{\partial}{\partial r}\right) \bar{F} = z^3 g / \bar{F}^2 \qquad (15)$$

Write now
$$\left.\begin{array}{c} x = \ln z = \ln r / y \\ y = \ln r \end{array}\right\} \qquad (16)$$

and put
$$1/y = \text{function of } (y-x) = w(s) \qquad (17)$$

where $s = y - x = \ln y \qquad (18)$

Then $\dfrac{d\ln w}{ds} = \dfrac{y}{w} \dfrac{dw}{dy} = -\dfrac{y \cdot \ddot{y}}{\dot{y}^2} \qquad (19)$

Our equations are now

$$H = \bar{F} - \dfrac{\partial \bar{F}}{\partial x} \qquad (20)$$

$$\left(\dfrac{\partial}{\partial x} + \dfrac{\partial}{\partial y}\right) P = \dfrac{e^{3x}}{\bar{F}^2}\left[\dfrac{\partial H}{\partial x} + \left(\dfrac{d\ln w}{ds}\right) H\right] \qquad (21)$$

$$\left(\dfrac{\partial}{\partial x} + \dfrac{\partial}{\partial y}\right) \bar{F} = g e^{3x}/\bar{F}^2 \qquad (22)$$

Boundary conditions

At origin  $R = \dot{R} = 0$

or  $R = 0$, $\dfrac{\partial R}{\partial x} = 0$ when $y = 0 \qquad (23)$

At shock-wave

when $x = 0$, $z = 1$, $F = 1 \qquad (24)$

(6) becomes $\dot{y}^2 P = \dot{y} \cdot \dot{y} H$

or  $P = H \qquad (25)$

(7) becomes $\dot{y}/g = \dot{y} - \dot{y}H$

$$\text{or} \quad g = 1 - H \tag{26}$$

and (8) becomes $\dfrac{2E}{\dot{y}^2 P} = 1/g - 1$

or $2E = \dot{y}^2 (1-g) P$

so

$$(1-g) P = H^2 = 2 w^2 E \tag{27}$$

# THE BRITISH SUPPLY COUNCIL IN NORTH AMERICA

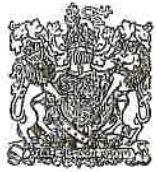

TELEPHONE
HAnover 2-2460

P. O. Box 51
WALL STREET STATION
NEW YORK 5, N. Y.

21st March, 1944

Dr. J.R. Oppenheimer,
P.O. Box 1539,
Santa Fe, New Mexico.

Dear Oppenheimer,

I enclose a copy of some notes by Davison on the boundary condition in the case of 2 media of different mean free path in case they are of interest to your group. I ought to explain that this is unfinished work - Davison abandoned this work temporarily at the time I left England and I took with me some manuscript notes of his which have been typed out here. You will see that he does not arrive at any very definite answer, but nevertheless the methods he uses may be of interest.

Yours sincerely,

R. Peierls

Dr. R. Peierls

RP:rg
Enc.
cc. Dr. Webster

# THE BRITISH SUPPLY COUNCIL IN NORTH AMERICA

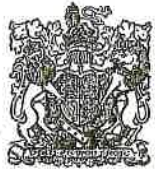

TELEPHONE
HAnover 2-2460

P. O. Box 51
WALL STREET STATION
NEW YORK 5, N. Y.

24th March, 1944.

Dr. J.R. Oppenheimer,
P.O. Box 1539,
Santa Fe, New Mexico.

Dear Oppenheimer,

I enclose a ~~reply~~ report by Wilson on the question of nearly spherical bodies which he wrote up at my request, following my conversation with Bethe on this point. I see that he has used diffusion theory. I believe it is not surprising that with the "elementary boundary condition" he obtains a very poor result because the effect will largely depend on the undisturbed density near the surface, which with the elementary boundary condition is zero.

I had hoped Wilson's calculations would include a general formula based on perturbation theory applied directly to the integral equation, which would be rather similar to Dirac's perturbation method. It is a fairly easy matter to do so, but I am reluctant to go into the details of this as this would involve first looking into the point that Bethe made, namely, that Dirac's method is incomplete and has to be supplemented by further terms. This no doubt would apply also to the present case.

The enclosed copy of Wilson's report is our only copy and I would be grateful if you could have it returned to me when in due course you receive an official copy from Washington.

I find that we did not at first obey the letter of the regulations by which receipts were to be sent with all SECRET letters. I would therefore be grateful if some time you could check that you have received my letters dated 15th February, 21st February, 9th March, which were sent without receipts.

Yours sincerely,

R. Peierls

Dr. R. Peierls

RP:rg
Enc.
cc: Dr. Webster

# THE BRITISH SUPPLY COUNCIL IN NORTH AMERICA

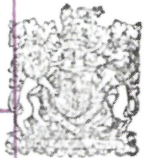

P. O. Box 51
Wall Street Station
New York 5, N. Y.

28th March, 1944

Dr. J. von Neumann
Institute of Advanced Studies
Princeton, N.J.

Dear von Neumann,

Following our recent conversation I am writing to give you the procedure used in our case to solve the boundary conditions in the case of a shock wave advancing into a stationary gas. I believe the method can be applied without much trouble to the more general case.

I shall assume, for simplicity, that r, t are the independent variables, and that the intervals chosen are equidistant both in space and time. Further I shall assume that, at a given time, the position of the shock wave coincides with the end of one of the intervals which we use. Then, at that instant, we can describe our state of knowledge by diagram 1. The variables known at each point from previous integration are indicated. R is the actual position, p the pressure, $\alpha$ the velocity. Note that $\alpha$ at $r = m$, $t = n + 1$ cannot be obtained directly because of the discontinuity.

Now assume a value for the position of the shock wave at $t = n + 2$, say $y_{n+2}$. Then, since the medium to the right is undisturbed, we have $R = y_{n+2}$ for $t = n + 2$, $r = y_{n+2}$. This is related to the shock wave velocity $V_{n+1}$ by

$$2 \epsilon V_{n+1} = y_{n+2} - y_n \qquad (1)$$

correct to second order inclusive, $\epsilon$ being the time interval. Next we can find the density at a point half-way between $r_{m-2}$ and $y_{n+2}$ which is denoted by $r^*$ in diagram 2:

$$r^* = \tfrac{1}{2}(y_{n+2} + r_{m-2}) \qquad (2)$$

The density is given by

$$\frac{\rho_0}{\rho^*} = \frac{R(r^*)^3 - y_{n+2}^3}{(r^*)^3 - y_{n+2}^3}$$

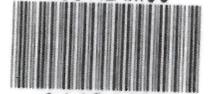

where $\rho_0$ is atmospheric density. Equation (3) is again correct up to terms of second order in the numerator.

We compare this with the density behind the shock-wave at $t = n + 1$ which is $\rho(n + 1, y_{n+1})$. The difference between this and $\rho^*$ is a small correction of first order and need therefore be known only to first order of accuracy. Now note that the line from $(n + 2, m - 3)$ to $(n, m-1)$ is parallel to, and twice as long as, the line from $(n + 2, r^*)$ to $(n + 1, y_{n+1})$ hence to first order we can write

$$\rho^* - \rho(n+1, y_{n+1}) = \frac{1}{2}\left[\rho(n+2, m-3) - \rho(n, m-1)\right] \quad (4)$$

The accuracy of this can, if necessary, be improved by computing also $\rho(n + 2, m - 5) - \rho(n, m-3)$ etc. and extrapolating towards the shock wave.

The density at the shock wave $\rho(n + 1, y_{n+1})$ is a function of $V_{n+1}$ by the usual boundary condition. Since by means of (1) (3) and (4) we can express this density as a function of $V_{N+1}$ we obtain an equation which can be solved for $y_{N+1}$.

**DOE ARCHIVES**

Having obtained $V_{N+1}$, and hence $y_{n+1}$, we can also obtain u and p at the shock wave. By interpolation we obtain $R(m, n + 2)$ and $u(m, n + 1)$ which are needed for continuing the integration process. From the value of p at the boundary we can obtain the entropy and hence define the adiabatic which this point will follow in future. By interpolation we obtain the entropy for the point $m - 1$. The interpolation requires the knowledge of the actual position $y_{n+1}$ of the shock wave at $t = n + 1$. This can be obtained with sufficient accuracy from values for previous times, together with the knowledge of $V_{n+1}$.

The process then repeats, except that the shock wave will not now coincide with the end of an interval and hence some slight modification of the method of extrapolation has to be used.

If the shock wave advances into a medium which is itself in motion, we have to know the values of the variables u, p, $\rho$ ahead of the shock front, in order to define the boundary conditions. It is probable that no trouble will be caused if one obtains these quantities by extrapolation, refining the process where necessary. For example, it is quite convenient to get the density ahead of the shock wave by substantially the same process as described above for the other side.

It is necessary for this approximation to have some knowledge of the point $y_{n+1}$ since one has to extrapolate to this point, but probably this point can be predicted with sufficient accuracy beforehand. Otherwise a correction may have to be applied after the point has been found.

I am afraid these notes are very sketchy, but my main object is to show that, if one can afford to take reasonably small intervals so that terms of higher order are indeed small, there is

nothing very terrible about having to introduce a discontinuity.

Since this can be done for the correct thermodynamic properties of the medium, I think it has an advantage over the process that you described. One question on that process. Behind the shock front, this will involve fairly high-period oscillations. In order to avoid appreciable errors, one must therefore use very small time intervals. Did you find this in practice an inconvenience, and how would the loss of time caused by this compare with the gain owing to avoiding the discontinuity?

Incidentally, have you thought at all about the following alternative way of avoiding discontinuity. In actual fact, the shock front has a finite width because of the viscosity and thermal conductivity of the medium. But artificially, assuming a viscosity very much larger than the actual, you can obtain instead of the discontinuity a front of a finite width, and as long as this width is small compared to the scale of the phenomenon, this should otherwise give no trouble. You pay for this by increasing the order of the equation as regards space derivatives, but the higher derivatives will be negligible except near the shock front. This is a macroscopic way of avoiding the discontinuity as opposed to your atomistic one. It is more complicated than yours but you can put in arbitrary heats and equation of state for the medium.

Yours sincerely,

Dr. R. Peierls

RP:rg
cc: Dr. Oppenheimer

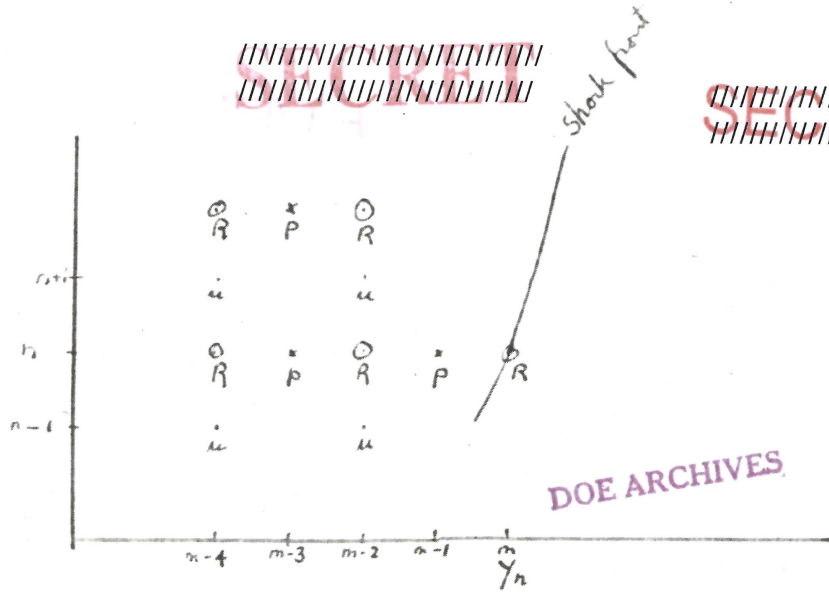

diagram 1

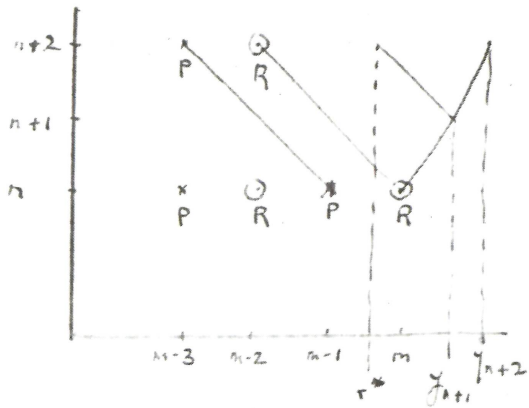

diagram 2

 

\end{document}